\documentclass[12pt, onecolumn,draftcls]{IEEEtran}
\usepackage{cite}
\usepackage{latexsym}
\usepackage{epsfig}
\usepackage{verbatim}
\usepackage{amssymb}

\usepackage{graphicx}
\usepackage{color}
\usepackage{epsfig}
\usepackage[cmex10]{amsmath}
\usepackage{stfloats}
\usepackage{setspace}
\usepackage{enumerate}
\usepackage{amsmath,amssymb,bm,amsfonts,bbm}
\usepackage{graphics,graphicx,epsfig,subfigure}
\usepackage{algorithm,algorithmic}
\usepackage{theorem}
\usepackage{threeparttable}
\usepackage{stfloats}
\usepackage{multirow}
\usepackage{mathrsfs}
\usepackage{color}
\usepackage{multirow}
\usepackage{subeqnarray}
\usepackage{cases}
\usepackage{wasysym}
\usepackage{flushend}

\newtheorem{mytheorem}{Theorem}
\newtheorem{mylemma}{Lemma}

\allowdisplaybreaks

\begin{document}
\title{\huge RF-based Energy Harvesting in Decode-and-Forward \\Relaying Systems: Ergodic and Outage Capacities}
\author{Yanju Gu,  and Sonia A\"issa
%\thanks{Manuscript received September 23, 2014; revised March 19, 2015 and May 24, 2015; accepted June 25, 2015. The editor coordinating the review of this manuscript and approving it for publication was Prof. Eduard Jorswieck. This work was supported by a Discovery Grant from the Natural Sciences and Engineering Research Council (NSERC) of Canada.}
%\thanks{Y. Gu and S. A\"issa are with the Institut National de la Recherche Scientifique (INRS-EMT), University of Quebec, Montreal, QC, H5A 1K6, Canada; Email: \{yanju.gu, sonia.aissa\}@emt.inrs.ca.}
%\thanks{%
%Digital Object Identifier ***}
}
\maketitle

%\markboth{IEEE Transactions on Wireless Communications, accepted for publication, 2015} {Gu \MakeLowercase{and} Aissa: RF-based Energy Harvesting in Decode-and-Forward Relaying Systems}

%%%%%%%%%%%%%%%%%%%%%%%%%%%%%%%%%%%
% Abstract
%%%%%%%%%%%%%%%%%%%%%%%%%%%%%%%%%%%

\begin{abstract}
\noindent Radio-frequency energy harvesting constitutes an effective way to prolong the lifetime of wireless networks, wean communication devices off the battery and power line, benefit the energy saving and lower the carbon footprint of wireless communications. In this paper, an interference aided energy harvesting scheme is proposed for cooperative relaying systems, where energy-constrained relays harvest energy from the received information signal and co-channel interference signals, and then use that harvested energy to forward the correctly decoded signal to the destination. The time-switching scheme (TS), in which the receiver switches between decoding information and harvesting energy, as well as the power-splitting scheme (PS), where a portion of the received power is used for energy harvesting and the remaining power is utilized for information processing, are adopted separately. Applying the proposed energy harvesting approach to a decode-and-forward relaying system with the three-terminal model, the analytical expressions of the ergodic capacity and the outage capacity are derived, and the corresponding achievable throughputs are determined. Comparative results are provided and show that PS is superior to TS at high signal-to-noise ratio (SNR) in terms of throughput, while at low SNR, TS outperforms PS. Furthermore, considering different interference power distributions with equal aggregate interference power at the relay, the corresponding system capacity relationship, i.e., the ordering of capacities, is obtained.
\end{abstract}
%
%\begin{keywords}
%\noindent Co-channel interference, decode-and-forward, energy harvesting, ergodic capacity, outage capacity, relaying, throughput.
%\end{keywords}
%%%%%%%%%%%%%%%%%%%%%%%%%%%%%%%%%%%
% INTRODUCTION
%%%%%%%%%%%%%%%%%%%%%%%%%%%%%%%%%%%
\section{Introduction}
In modern society, wireless communication devices are omnipresent and have become numerous.
They are intensively involved in different applications such as video and audio information transmission \cite{WSN}, monitoring in modern healthcare systems \cite{Monitor} and safety message exchange in vehicular networks \cite{Proceedings}.
On one hand, the energy consumption is tremendous \cite{Green}, which makes energy saving for communication a critical problem to be solved. On the other hand, recharging by traditional wiring method or battery replacement is not feasible for such huge number of small devices, like sensors, implantable medical devices, etc.
Although far-field microwave power transfer is a strong candidate to replace cables in long-distance power transfer, additional power beacons need to be settled and deployed, which is not a ready work for nowadays communication systems \cite{HuangTWC,XiaICC,XiaICC2,XiaTSP}.

Energy harvesting has become an appealing solution to such problems \cite{EH1,EH5,wangpower}.
Energy from solar, vibration, thermoelectric effects, and so forth \cite{EH2}, can be harvested and converted to electrical energy to support these energy-constrained communication devices \cite{Harvest-Use2}.
%Common energy harvesting devices are solar cells, wind turbines and piezoelectric cells.
A promising harvesting technology is to use the radio frequency (RF) energy, since ambient RF signals, e.g., from TV broadcast and cellular communications, are widely available in urban areas (day and night, indoors and outdoors) \cite{SIGCOMMbestpaper}.
In this technique, the ambient RF radiation is captured by the receive antennas of wireless devices and converted into direct current voltage through appropriate circuits \cite{EH3, EH4}. A safe way has been advanced to wirelessly power chips in human body by using such method \cite{Ada}.

As the signal carries information as well as energy at the same time, simultaneous wireless information and power transfer has been studied recently, where the receiver is assumed to be able to decode the information and harvest energy from the same signal \cite{SEH1, SEH2}.
However, due to practical circuit limitations, it is difficult to harvest the energy and decode the information at the same time.
There are two schemes for harvesting energy and decoding information separately \cite{Neg,TSplit,Zhou_AF,zhang2}, one is the time-switching scheme (TS) in which the receiver switches over time between decoding information and harvesting energy;
and the other is the power-splitting scheme (PS) in which a portion of the received power is used for energy harvesting and the remaining power is utilized for the information processing.
From the perspective of receiver's complexity, TS is superior to PS in that commercially available circuits that are separately designed for information decoding and energy harvesting can be used.

Simultaneous information decoding and energy harvesting has applications and advantages in wireless systems in general, whether in point-to-point communication or when nodes cooperate together in delivering the source signal to its final destination, \cite{dingzhiguo,zhang2}.
Indeed, in cooperative networks, by deploying relays between the source and the destination, the cover range and capacity of the communication system can be enhanced.
However, the relays may have limited battery and wired charging may be difficult to be implemented when and where needed. To prolong the lifetime of relaying systems, wireless energy harvesting at the relays becomes a necessity \cite{GuICC14,Zhou_AF,R2,R1}.

Since the radio signal propagates freely over space, a receiver would receive the desired signal with a superposition of unwanted signals, namely interferences, which in turn results in low capacity between the transmitter and the receiver.
Interference is the primary bottleneck on the data rate capacity of most wireless networks.
How to decrease or avoid interference  and increase the signal-to-interference-plus-noise ratio (SINR) has always been a big concern in research and industry.
Techniques such as frequency reuse \cite{FreReUse}, multi-cell coordination \cite{Multicell} and interference alignment \cite{InterferAlignment,YangGlobecom} have been proposed for interference cancellation.
While interference decreases the communication system capacity, from the energy  point-of-view the interference signal provides additional energy for the harvesting system.
Therefore, investigating the role that the interference plays in energy-harvesting based communication system is of major importance, though still missing.

In this paper, a decode-and-forward (DF) relaying system where the relays need to replenish energy from the received RF signals, is considered.
For the limitation of hardware, harvest-use strategy in which no device equipment is dedicated to store the harvested energy, is adopted \cite{Harvest-Use2}.
As opposed to traditional relaying where co-channel interference (CCI) within the same bandwidth as the transmitted signal deteriorates the system performance \cite{Xia-Tcom12,Gu}, and have to be eliminated by applying interference alignment approach or by decoding the interfering signals when they are strong, in this work, CCI signals are utilized as a new source of power for relay recharging. Specifically, the relays harvest energy from both the information signal and the CCI signals, and then use that energy to decode the source signal and forward it to the destination node. In this way, the interference acts as useful power in the energy harvesting phase and as noise in the information decoding phase. Initial results for the ergodic capacity of a DF relaying system with TS protocol appear in \cite{GuICC14}.

To provide a thorough study and guidelines for practical applications, both TS and PS energy harvesting schemes are investigated.
The analysis of the system performance is challenging due to the random feature of the transmission power at the relay in the proposed energy harvesting system.
First, the ergodic capacity, which is a fundamental performance indicator for delay-insensitive services, when the codeword length can be sufficiently long to span over all the fading blocks, is investigated.
Moreover, for real-time applications, a more appropriate performance metric, the outage capacity, defined as the maximum constant rate that can be maintained over fading blocks with a given outage probability, is studied.
Analytical expressions for both the ergodic capacity and the outage capacity are derived and the corresponding achievable throughputs are obtained.
In addition, the impacts of the interference power distribution on the ergodic capacity and the outage capacity as well as the corresponding achievable throughputs of the proposed energy harvesting system are also studied based on the majorization theory.

The rest of this paper is organized as follows. The system model and energy harvesting schemes are described in Section II. Considering time-switching and power-splitting protocols, the ergodic capacity and outage capacity are analyzed in Section III and Section IV, respectively. Simulation results which corroborate the analytical results are provided in Section V. Finally, the paper's conclusion is presented in Section VI.
%
%%%%%%%%%%%%%%%%%%%%%%%%%%%%%%%%%%%
\section{Energy-Harvesting Based Relaying}\label{Model}
\subsection{System and Channel Models}
We consider a cooperative DF relaying system, where the source $S$ communicates with the destination $D$ through the help of an energy-constrained intermediate relaying node $R$, as shown in Fig. 1 (a).
Each node is equipped with a single antenna and operates in the half-duplex mode in which the node cannot simultaneously transmit and receive signals in the same frequency band.
Both, the first hop (source-to-relay) and the second hop (relay-to-destination), experience independent Rayleigh fading with the complex channel fading gains given by $h \sim CN(0,\Omega_{h})$ and $g \sim CN(0,\Omega_{g})$, respectively. The channels follow the block-fading model in which the channel remains constant during the transmission of a block and varies independently from one block to another.
The channel state information is only available at the receiver.

Wireless communication networks are generally subjected to CCI due to the aggressive frequency reuse for a more efficient resource utilization. In this vein, we assume that there are $M$ CCI signals affecting the  relay.
The CCI signals are assumed independent but not identically distributed.
Specifically, the channel fading gain between the $i^{\rm th}$ interferer and the relay node $R$, denoted  $\beta_i$, is modeled as $\beta_i\sim CN(0,\Omega_{\beta_i})$. Hereafter, the desired channels and the interference channels are assumed to be independent from each other.
\begin{figure}[!ht]
  \centering
\vspace*{-0.05in}\mbox{\subfigure[]{\epsfig{figure=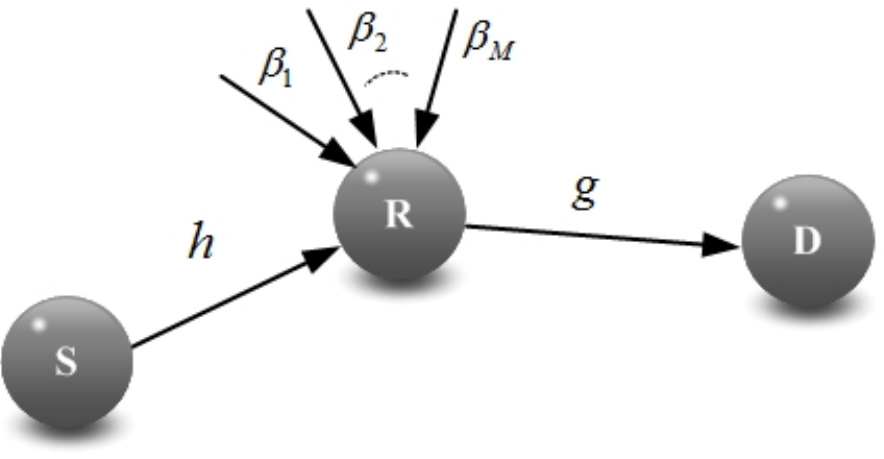,width=2.in}}\label{DF} }\\
\vspace*{-0.05in}\mbox{\subfigure[]{\epsfig{figure=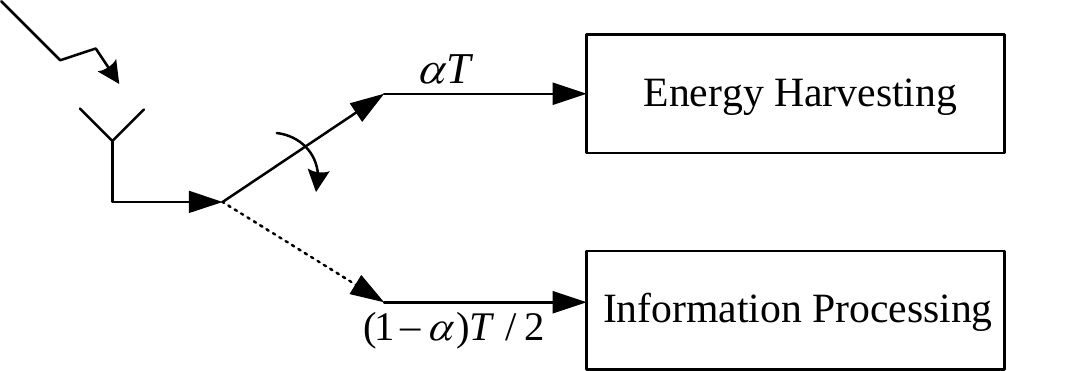,width=2.5in}}\label{timesplit} }\\
\vspace*{-0.05in}\mbox{\subfigure[]{\epsfig{figure=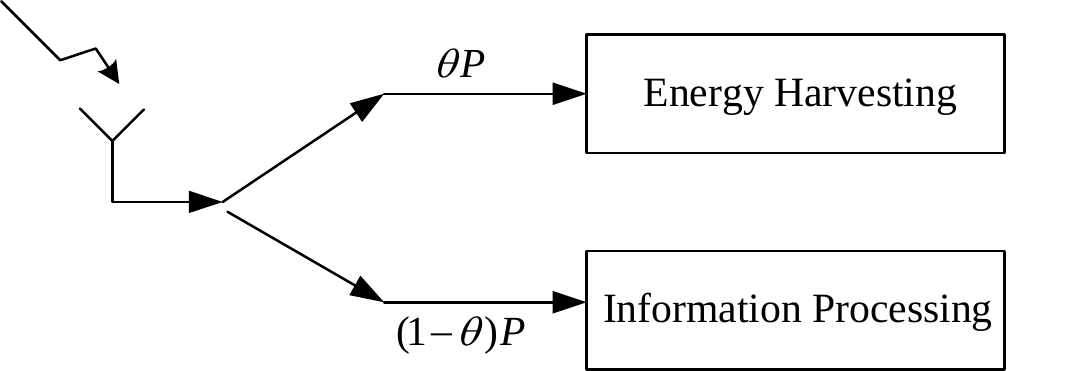,width=2.5in}}\label{powersplit} }\\
\caption{(a) Energy-harvesting based relaying system, where the energy-constrained relay harvests energy from the signal and the co-channel interferences. (b) Time-switching protocol for energy harvesting and information processing at the relay. (c) Power-splitting protocol for energy harvesting and information processing at the relay.}
\label{SystemModel}
\end{figure}
\subsection{Wireless Energy Harvesting at the Relay}
In the network under study, the relay is considered to be constrained in terms of energy. That is, it may have limited battery reserves and needs to rely on some external charging mechanism in order to remain active in the network, and assist the communication process between the source, $S$, and the destination, $D$, as required. In the proposed approaches, the received interference and information signals at the relay are exploited to replenish energy for the relay. Both TS and PS architectures for harvesting energy are studied.

\noindent\textit{1) Time-Switching Scheme:}

The time-switching based protocol is adopted at the relay node as illustrated in Fig. 1(b), where $T$ is the block time in which a certain block of information is transmitted from the source node to the destination node and $\alpha$, with $0\leq\alpha\leq1$, is the fraction of the block time in which the relay harvests energy from the received interference signal and information signal. The remaining block time is divided into two equal parts, namely $(1-\alpha)T/2$, for information transmission from the source to the relay and from the relay to the destination, respectively.
Since there is no energy buffer to store the harvested energy (Harvest-Use) \cite{Harvest-Use1,Harvest-Use2}, all the energy collected during the harvesting phase is consumed by the relay.

In the first-hop phase, source $S$ transmits signal $s$ with power $P_{_S}$ to the relay $R$. Accordingly, the received signal at the relay is given by
\begin{equation} \label{SR}
y_{_{SR}} = \sqrt {P_{_S}}hs + \sum_{i = 1}^M \sqrt{P_i} \beta_is_i + n_{_R},
\end{equation}
where $s_i$ and $P_i$ denote the signal and its corresponding power, from the $i^{\rm th}$ interferer, and $n_{_R}$ is
the additive white Gaussian noise (AWGN) at the relay with zero mean and variance $\sigma^2_R$.

Accordingly, the received SINR at the relay is given by
\begin{eqnarray}
\label{SINR-R}
\gamma_{_{SR}} &=&
\displaystyle\frac{P_{_S}|h|^2}{\sigma_{_R}^2 + \sum_{i = 1}^{M} P_i|\beta_i|^2} \nonumber \\
&=& \displaystyle\frac{\gamma_{h}}{1 + I_R},
\end{eqnarray}
%\begin{equation}
%\label{SINR-R}
%\gamma_{_{SR}} =
%\displaystyle\frac{P_{_S}|h|^2}{\sigma_{_R}^2 + \sum_{i = 1}^{M} P_i|\beta_i|^2} =\displaystyle\frac{\gamma_{h}}{1 + I_R},
%\end{equation}
where $\gamma_{h} \triangleq  \frac{P_{_S}}{\sigma_{_R}^2}|h|^2$ and $I_R \triangleq \sum_{i=1}^{M}\frac{P_i}{\sigma^2_R}|\beta_i|^2$.
The received data is correctly decoded if the instantaneous received SINR $\gamma_{_{SR}}$ at the relay is higher than the pre-defined threshold $\gamma_{_\mathrm{th}}$.

When the relay is active, it harvests energy from the received information signal and the interference signal for a duration of $\alpha T$ at each block, and thus, the harvested energy is obtained as
\begin{equation} \label{Eh}
E_{_{H}} = \eta\bigg(P_{_S} |h|^2 + \sum_{i = 1}^{M}P_i|\beta_i|^2\bigg)\alpha T,
\end{equation}
where
$\eta$ is the energy conversion efficiency coefficient, with value varying from $0$ to $1$ depending upon the harvesting circuitry.

Since the processing power required by the transmit/receive circuitry at the relay is generally negligible compared to the power used for signal transmission \cite{TSplit, Neg}, here we suppose that all the energy harvested from the received signals (the source's and the CCI's)  is consumed by the relay for forwarding the information to the destination.
Therefore, from (\ref{Eh}), the transmission power of the relay is readily given by
%\begin{equation}\label{P-TS}
%P_{_{R}} = \displaystyle\frac{E_{_{H}}}{(1 - \alpha)T/2} = \displaystyle\frac{2\alpha \eta\sigma_R^2}{1 - \alpha } (\gamma_{h} + I_R).
%\end{equation}
\begin{eqnarray}\label{P-TS}
P_{_{R}} &=& \displaystyle\frac{E_{_{H}}}{(1 - \alpha)T/2} \nonumber \\
&=& \displaystyle\frac{2\alpha \eta\sigma_R^2}{1 - \alpha } (\gamma_{h} + I_R).
\end{eqnarray}
Then, the received signal at the destination node $D$ is given by
\begin{equation} \label{y-D}
y_{_{RD}} = \sqrt {P_{_R}} g s_{_R} + n_{_D},
\end{equation}
where $s_{_R}$ is the signal transmitted from the relay and $n_{_D}$ is the AWGN noise at the destination, with zero mean and variance $\sigma_D^2$. From (\ref{y-D}), the received signal-to-noise ratio (SNR) at the destination node is obtained as
%\begin{equation} \label{SNR-D}
%\gamma_{_{RD}} = \displaystyle\frac{P_{_{R}}|g|^2}{\sigma_{_D}^2} = \underbrace{\displaystyle \frac{2\alpha\eta}{1-\alpha} \frac{\sigma_R^2}{\sigma_D^2} |g|^2}_{\triangleq W}(\gamma_{h} + I_R),
%\end{equation}
\begin{eqnarray} \label{SNR-D}
\gamma_{_{RD}} &=& \displaystyle\frac{P_{_{R}}|g|^2}{\sigma_{_D}^2} \nonumber \\
&=& \underbrace{\displaystyle \frac{2\alpha\eta}{1-\alpha} \frac{\sigma_R^2}{\sigma_D^2} |g|^2}_{\triangleq W}(\gamma_{h} + I_R),
\end{eqnarray}
where the defined random variable $W$ follows the same distribution as of $|g|^2$.

\textit{\noindent 2) Power-Splitting Scheme:}

In this case, the protocol adopted at the relay node is as illustrated in Fig. 1(c), where $P$ is the power of the received signal and $\theta$, with $0\leq\theta\leq1$, is the fraction of power that the relay harvests from the received interference and information signal. The remaining power is $(1-\theta)P$, which is used for information detection. In this paper, we consider a pessimistic case in which power splitting only reduces the signal power, but not the noise power, which can provide a lower-bound performance measure for relaying networks in practice.

Accordingly, after power-splitting, the received signal at the relay for information detection is given by
\begin{equation} \label{Signal.P}
y_{_{SR}} = \sqrt {(1-\theta)P_{_S}}hs + \sum_{i = 1}^{M} \sqrt{(1-\theta)P_i} \beta_i s_{i} + n_{_{R}}.
\end{equation}
Then, the received SINR at the relay is obtained as
\begin{eqnarray}
\label{SINR.P}
\gamma_{_{SR}} &=& \displaystyle\frac{(1-\theta)P_{_S}|h|^2}{\sigma_R^2 + \sum_{i = 1}^{M} (1-\theta)P_i|\beta_i|^2} \nonumber \\
&=& \displaystyle\frac{\gamma_{h}}{1 + I_R},
\end{eqnarray}
%\begin{equation}
%\label{SINR.P}
%\gamma_{_{SR}} = \displaystyle\frac{(1-\theta)P_{_S}|h|^2}{\sigma_R^2 + \sum_{i = 1}^{M} (1-\theta)P_i|\beta_i|^2} = \displaystyle\frac{\gamma_{h}}{1 + I_R},
%\end{equation}
where ${\gamma}_h \triangleq\frac{(1-\theta)P_{_S}}{\sigma_{_R}^2}|h|^2$ and ${I}_R \triangleq \sum_{i=1}^{M}\frac{(1-\theta)P_i}{\sigma^2_R}|\beta_i|^2$.
Note that  ${\gamma}_h $ and ${I}_R$ here have distinct denotations from those in (\ref{SINR-R}) for the TS protocol. We use the same symbols ${\gamma}_h $ and ${I}_R$ in order to unify the analysis in the following sections.

Different from TS, for PS, the relay harvests energy from the received information and interference signal for a duration of $T/2$ at each block, and thus, the harvested energy at the relay is obtained as
\begin{equation} \label{Harvest.P}
E_{_{H}}= \eta{\theta}\left(P_{_S} |h|^2 + \sum_{i = 1}^{M}P_i|\beta_i|^2\right)\frac{T}{2}.
\end{equation}
Suppose that all the harvested energy is consumed by the relay for forwarding the information to the destination node in the second-hop phase.
From (\ref{Harvest.P}), the transmission power of the relay node is readily given by
%\begin{equation}\label{P-PS}
%P_{_{R}} = \frac{E_{_{H}}}{T/2} = \frac{{\eta \theta \sigma _R^2}}{1 - \theta }
%\left({\gamma}_h+{I}_R\right).
%\end{equation}
\begin{eqnarray}\label{P-PS}
P_{_{R}} &=& \frac{E_{_{H}}}{T/2} \nonumber \\
&=& \frac{{\eta \theta \sigma _R^2}}{1 - \theta }
\left({\gamma}_h+{I}_R\right).
\end{eqnarray}
Then, the received SNR at the destination node is expressed as
%\begin{equation}\label{SNRD}
%\gamma_{_{RD}} = \frac{P_R| g |^2}{\sigma _D^2} = \underbrace{\frac{\eta \theta }{( 1 - \theta) } \frac{\sigma_R^2}{ \sigma _D^2 }{|g|^2}}_{\triangleq W}({\gamma}_h+{I}_R).
%\end{equation}
\begin{eqnarray}\label{SNRD}
\gamma_{_{RD}} &=& \frac{P_R| g |^2}{\sigma _D^2} \nonumber \\
&=& \underbrace{\frac{\eta \theta }{( 1 - \theta) } \frac{\sigma_R^2}{ \sigma _D^2 }{|g|^2}}_{\triangleq W}({\gamma}_h+{I}_R).
\end{eqnarray}

By introducing the random variables ${\gamma}_h $, ${I}_R$ and $W$, we unify the derivations of the distribution of the end-to-end SNR and the capacity metrics for these two schemes (TS and PS), as detailed in the following sections.
\section{Ergodic Capacity}\label{Ergodic Capacity }
In this section, the exact closed-form cumulative distribution function (CDF) of the end-to-end SNR is derived. Then, the ergodic capacity and the corresponding achievable throughput are investigated for the energy harvesting DF relaying system with time-switching or power-splitting. The impact of the interference power distribution on the ergodic capacity and achievable throughput is also analyzed, based on the majorization theory.
\subsection{End-to-End SNR}
All channels,  i.e., $h$, $\{\beta_i\}^M_{i=1}$ and $g$ are supposed to be subject to independent Rayleigh fading.
Then, the received SNR at the first hop, $\gamma_h$, is of exponential distribution with the probability density function (PDF) given by
\begin{equation}\label{pdfX}
f_{\gamma_{h}}( x ) = \frac{1}{\bar \gamma_h}\exp \left( { - \frac{x}{{{{\bar \gamma }_h}}}} \right),\;\;x \ge 0,
\end{equation}
where ${\bar\gamma}_h $, equal to $\frac{P_{_S}}{\sigma_R^2}\Omega_h$ for TS and to $\frac{(1-\theta)P_{_S}}{\sigma_R^2}\Omega_h$ for PS, is the average SNR from the source to the relay in a given time slot.

The quantity $I_R$ is the sum of $M$ statistically independent and not necessarily identically distributed (i.n.i.d.) exponential random variables, each with mean $\mu_i = \frac{P_i}{\sigma_R^2}\Omega_{\beta_i}$ for the TS-based scheme and $\mu_i = \frac{(1-\theta)P_i}{\sigma_R^2}\Omega_{\beta_i}$ for the PS-based one.
Thus, the PDF of $I_R$ can be explicitly obtained as
\begin{equation} \label{pdfY}
f_{I_R}(y)
=  \sum_{i = 1}^{\upsilon ( \bm {A} )} \sum_{j = 1}^{\tau_i(\bm {A})} \chi_{i,j}(\bm {A})
\frac{\mu_{\langle i \rangle }^{- j}}{( j - 1 )!} y^{j - 1}
\exp \bigg(-\frac{y}{\mu_{\langle i \rangle }}\bigg),\;\;y \ge 0,
\end{equation}
where matrix $\bm {A} = \mathrm{diag}(\mu_1, \mu_2, \ldots, \mu_{_M})$, $\upsilon ( \bm {A} )$ denotes the number of distinct diagonal elements of $\bm {A}$, $\mu_{\langle 1 \rangle } > \mu_{\langle 2 \rangle } > \ldots > \mu_{\langle \upsilon ( \bm {A} ) \rangle}$ are the distinct diagonal elements in decreasing order, $\tau_i(\bm {A})$ is the multiplicity of $\mu_{\langle i \rangle }$, and $\chi_{i,j}(\bm {A})$ is the $(i, j)$th characteristic coefficient of $\bm {A}$ \cite{Win2}. Note that when the interfering signals are statistically independent and identically distributed (i.i.d.), i.e., $\mu_i = \mu$ for $i=1, 2, \ldots, M$, then $\upsilon ( \bm {A} )=1$, $\tau_1(\bm {A})=M$ and $I_{_R}$ is a sum of $M$ i.i.d. exponential random variables with the central chi-squared distribution given by $f_{I_{_R}}(\gamma)
=\frac{\mu^{- M}}{( M - 1 )!} \gamma^{M - 1}\exp \left(-\frac{\gamma}{\mu}\right)$.

The CDF of $\gamma _{_{SR}}$ is then obtained as
\begin{eqnarray} \label{F_SR}
F_{\gamma_{_{SR}}}(\gamma)
 %&=& \Pr \left\{ {\frac{{{\gamma _h}}}{{1 + {I_R}}} \le \gamma } \right\}\nonumber \\
 %&=& {\mathbb{E}_{{I_R}}}\left\{ {{F_{{\gamma_{_{SR}}}\left| {{I_R}} \right.}}\left( \gamma  \right)} \right\}\\
 &=& {\mathbb{E}_{{I_R}}}\left\{ {1 - \exp \left[ { - \frac{{\gamma \left( {1 + {I_R}} \right)}}{{{{\bar \gamma }_h}}}} \right]} \right\} \label{F_SR1}\\
 &=& 1- \exp \left(-\frac{\gamma}{{\bar \gamma}_h}\right)\sum_{i = 1}^{\upsilon ( \bm {A} )} \sum_{j = 1}^{\tau_i(\bm {A})} \chi_{i,j}(\bm {A})\nonumber \\
 & & \times\bigg( 1 + \frac{\mu_{\langle i\rangle}}{{\bar \gamma }_h}\gamma \bigg)^{ - j}, \label{F_SR2}
\end{eqnarray}
where $\mathbb{E}[\cdot]$ denotes the statistical expectation operator.
In the case when the interfering signals are i.i.d., the CDF of $\gamma _{_{SR}}$ reduces to
$F_{\gamma_{_{SR}}}(\gamma)
= 1- \left( 1 + \frac{\mu}{{\bar \gamma }_h}\gamma \right)^{ - M}
\exp \left(-\frac{\gamma}{{\bar \gamma}_h}\right)$.

Similar to the derivation of (\ref{F_SR2}), the PDF of $\gamma_{_{RD}}$, which involves products of two random variables, is determined as
\begin{eqnarray} \label{Eq.10}
F_{\gamma_{_{RD}}}(\gamma)
 &=& 1 - \sum_{i = 1}^{\upsilon (\bm {B})} \sum_{j = 1}^{\tau_i(\bm {B})} \frac{\chi_{i,j}(\bm {B})}{(j - 1)!}
2 \bigg(\frac{\gamma}{{\bar \gamma }_g\mu_{\langle i\rangle }} \bigg)^{\frac{j}{2}} \nonumber \\
 & & \times K_j\bigg(2\sqrt {\frac{\gamma}{{\bar \gamma }_g\mu_{\langle i \rangle}}} \bigg),
\end{eqnarray}
where ${\bar \gamma }_g = \frac{2\alpha \eta}{1 - \alpha }
\frac{\sigma_R^2}{\sigma_D^2}\Omega_g$ for TS and ${\bar \gamma }_g = \frac{\theta \eta}{1 - \theta }
\frac{\sigma_R^2}{\sigma_D^2}\Omega_g$ for PS, $\bm {B} = \mathrm{diag}(\mu_1, \mu_2, \ldots, \mu_{_{M+1}})$ with
$\mu_{_{M+1}}= {\bar \gamma }_h$, $\upsilon ( \bm {B} )$ denotes the number of distinct diagonal elements of $\bm {B}$, $\mu_{\langle 1 \rangle } > \mu_{\langle 2 \rangle } > \ldots > \mu_{\langle \upsilon ( \bm {B} ) \rangle}$ are the distinct diagonal elements in decreasing order, $\tau_i(\bm {B})$ is the multiplicity of $\mu_{\langle i \rangle }$, $\chi_{i,j}(\bm {B})$ is the $(i, j)$th characteristic coefficient of $\bm {B}$, and $K_{j}(\cdot)$ stands for the $j$th-order modified Bessel function of the second kind \cite{Gradshteyn}.
\subsection{Ergodic Capacity and Achievable Throughput}
Ergodic capacity, in the unit of bit/s/Hz, quantifies the ultimate reliable communication limit of the fading channel. It is only achievable with infinite coding delay.
Ergodic capacity can be obtained by averaging the instantaneous capacity over all fading states. In the DF-cooperative communication system under study, the instantaneous capacity is determined by the minimum one of each individual link, i.e., the first- and second-hop links. Therefore, the ergodic capacity is expressed as
\begin{eqnarray}
C_\mathrm{erg}\!\!\!\!
&=&\!\!\!\! \mathbb{E}\bigg[\textrm{min}\bigg \{\frac{1}{2}\log_2(1+\gamma_{_{SR}}), \frac{1}{2} \log_2(1+\gamma_{_{RD}})\bigg\}\bigg]
\label{capacity1a}\\
&=&\!\!\! \! \mathbb{E}\bigg[\frac{1}{2}\log_2\left(1+\min \{\gamma_{_{SR}},\gamma_{_{RD}}\}\right)\bigg] \label{capacity1b}\\
&=& \!\!\!\! \frac{1}{2}\int_0^{\infty}\log_2(1+\gamma)f_{\gamma_\textrm{min}}(\gamma)d\gamma,\label{capacity1c}
\end{eqnarray}
where $f_{\gamma_\textrm{min}}(\gamma)$ stands for the PDF of the random variable $\min \{\gamma_{_{SR}},\gamma_{_{RD}}\}$. The factor $1/2$ in (\ref{capacity1a}) is introduced by the fact that two transmission phases are involved in the system. Expression (\ref{capacity1b}) follows from the strictly monotonically increasing property of the logarithm function for non-negative real numbers.

Using the integration-by-parts method, (\ref{capacity1c}) can be rewritten as
\begin{eqnarray}
C_\mathrm{erg}\label{capacity2a}
\!\!\!\!&=&\!\!\!\!\! \frac{1}{2}\left\{\log_2(1 + \gamma )[ F_{\gamma_{\min}}( \gamma  ) - 1 ]\right\}_0^\infty \nonumber\\
&~& -\frac{1}{2\ln 2}\int_0^\infty \!\! \frac{1}{1 + \gamma }[ F_{\gamma_{\min}}( \gamma  ) - 1 ] d\gamma \\
&=&\!\!\!\!\! \frac{1}{2\ln 2}\int_0^\infty  \!\!\!\!\frac{1}{1 + \gamma }[1- F_{\gamma_{\min}}( \gamma  )] d\gamma,
\label{capacity2b}
\end{eqnarray}
where in (\ref{capacity2a}) the operator $\{f(x)\}_a^b \triangleq f(b)-f(a)$ and $F_{\gamma_\textrm{min}}(\gamma)$ denotes the CDF of the random variable $\min \{\gamma_{_{SR}},\gamma_{_{RD}}\}$ and is given by
\begin{eqnarray}\label{CDF_DM}
F_{\gamma_\textrm{min}}(\gamma)
\!\!\!\!\! &=& \!\!\!\!\! F_{\gamma_{_{SR}}}(\gamma)\!+\! F_{\gamma_{_{RD}}}(\gamma)\!-\!
 \Pr \left\{ {\gamma_{_{SR}}\!\!\leq\!\! \gamma, \gamma_{_{RD}}\!\!\leq\!\! \gamma } \right\} \\
\!\!\!\!\! &=& \!\!\!\!\! \Pr \left\{ {\gamma_{_{SR}} \leq \gamma } \right\}+ \Pr \left\{ {\gamma_{_{SR}} > \gamma, \gamma_{_{RD}} \leq \gamma } \right\}. \label{CDF_DM1}
\end{eqnarray}
If $\gamma$ is set to be a pre-defined threshold, (\ref{CDF_DM1}) is the expression of the outage probability at the destination, the detailed derivation of which is illustrated in the following section.

The achievable throughput at the destination relates only to the effective information transmission time and is given by
\begin{eqnarray}\label{throughput_TS}
T_\mathrm{erg}
 &=&  \frac{(1-\alpha)T}{T}C_\mathrm{erg} \nonumber \\
 &=& (1-\alpha)C_\mathrm{erg},
\end{eqnarray}
for the system with TS protocol, and by
\begin{equation}\label{throughput_PS}
T_\mathrm{erg} = C_\mathrm{erg},
\end{equation}
for the PS based scheme.

Different from the conventional relaying system with no rechargeable nodes, from (\ref{throughput_TS}) and (\ref{throughput_PS}) it is clear that in the interference aided energy harvesting system, the achievable throughput depends not only on $P_{_S}$, $\sigma_R^2$ and $\sigma_D^2$, but also on $\alpha$ or $\theta$, $\eta$ and $P_i$.
\subsection{Impact of Interference Power Distribution}
In order to provide an analysis of the impact of the interference power distribution on the energy harvesting system performance when the total received interference power is the same, in this section, the Schur-convex property of ergodic capacity and throughput is investigated.

For two vectors $\bm{x}$ and $\bm{y} (\in \mathbb{R}^{n}$) with descending ordered components $x_{1}\geq x_{2}\geq \cdots \geq x_{n}\geq 0$ and $y_{1}\geq y_{2}\geq \cdots \geq y_{n}\geq 0$, respectively, one can say that the vector $\bm {x}$ majorizes the vector $\bm {y}$ and writes
$\bm {x}\succeq \bm {y}$ if $\sum\limits_{k = 1}^m {{x_k}}  \ge \sum\limits_{k = 1}^m {{y_k}}$ for $m = 1, \ldots , n - 1$, and $\sum\limits_{k = 1}^n {{x_k}}  = \sum\limits_{k = 1}^n {{y_k}}$.
A real-valued function $\Phi$ defined on $\mathcal{A}\subset \mathbb{R}^{n}$ is said to be Schur-convex on $\mathcal{A}$ if $\bm {x}\succeq \bm {y}$ on $\mathcal{A} \Rightarrow \Phi(\bm {x})\geq \Phi(\bm {y})$.

Assume that $\omega _{1},\ldots ,\omega _{n}$ are i.i.d. random variables according to a given PDF. Furthermore, assume vector $ \bm{\mu} $ to have non-negative entries that are ordered in non-increasing order
$\mu_{1}\geq \mu_{2}\geq \cdots \geq \mu_{n}\geq 0$.

\begin{mylemma}\label{ExpectedSchur}
Suppose the function $f : \mathbb{R}_{+} \rightarrow \mathbb{R}_{+}$ is concave. Then,
\begin{equation}\label{WeightedSum}
G\left( \bm{\mu}  \right) = \mathbb{E}_{\omega _{1},\ldots ,\omega _{n}}\left[ {f\left( {\sum\limits_{k = 1}^n {{\mu _k}{\omega _k}} } \right)} \right]
\end{equation}
is Schur-concave. Assume $f$ is convex. Then the function $G$ in (\ref{WeightedSum}) is Schur-convex \cite{Majorization}.
\end{mylemma}

\begin{mytheorem}\label{Concave_Erg}
The ergodic capacity, $C_\mathrm{erg}$, and the achievable throughput, $T_\mathrm{erg}$, of the interference aided energy-harvesting DF relaying system are Schur-convex with respect to $\bm{\mu}$, where $\bm{\mu}=(\mu _{1},\mu _{2},\ldots ,\mu _{M})$ with $\mu_{1}\geq \mu_{2}\geq \cdots \geq \mu_{_M}\geq 0$.
\end{mytheorem}
\begin{IEEEproof}
We can see that ${I}_R =\sum_{i=1}^{M}\mu_{i}\omega_{i}$, where $\omega _{1},\ldots ,\omega _{n}$ are i.i.d. standard exponentially distributed with unit mean and $\sum_{i = 1}^{M} {\mu_{i}} = \mathbb{E}\{ {I}_R \}$.
According to (\ref{F_SR1}),
\begin{eqnarray}
F_{\gamma_{_{SR}}}(\bm{\mu})
&=& {\mathbb{E}_{{I_R}}}\left\{ {1 - \exp \left[ { - \frac{{\gamma \left( {1 + {I_R}} \right)}}{{{{\bar \gamma }_h}}}} \right]} \right\} \nonumber\\
&=& {\mathbb{E}_{{I_R}}}\left\{ f(I_R) \right\}.
\end{eqnarray}
Since the second derivative $f''\left( {{I_R}} \right) =  - \frac{{{\gamma ^2}}}{{{{\bar \gamma }_h}^2}}\exp \left[ { - \frac{{\gamma \left( {1 + {I_R}} \right)}}{{{{\bar \gamma }_h}}}} \right]\leq 0$, then $f(I_R)$ is a concave function. Thus, according to Lemma \ref{ExpectedSchur}, $F_{\gamma_{_{SR}}}(\bm{\mu})$ is Schur-concave with respect to $\bm{\mu}$.
Similarly, having
\begin{eqnarray}
F_{\gamma_{_{RD}}}(\bm{\mu})
&=& \mathbb{E}_{W, {I_R}}\left\{ {1 - \exp \left[ { - \frac{1}{{{{\bar \gamma }_h}}}\left( {\frac{\gamma }{W} - {I_R}} \right)} \right]} \right\} \nonumber\\
&=& \mathbb{E}_{W, {I_R}}\left\{  g(I_R) \right\},
\end{eqnarray}
and since $g''\left( {{I_R}} \right) =  - \frac{{1}}{{{{\bar \gamma }_h}^2}}\exp \left[ { - \frac{1}{{{{\bar \gamma }_h}}}\left( {\frac{\gamma }{W} - {I_R}} \right)} \right]\leq 0$, the function $F_{\gamma_{_{RD}}}(\bm{\mu})$ conditioned on $W$ is Schur-concave.
That is, for any two vectors $\bm{\mu}_1\succeq \bm{\mu}_2$, $F_{\gamma_{_{RD}}}(\bm{\mu}_1|W)\leq F_{\gamma_{_{RD}}}(\bm{\mu}_2|W)$. Averaging over $W$, we have $F_{\gamma_{_{RD}}}(\bm{\mu}_1)\leq F_{\gamma_{_{RD}}}(\bm{\mu}_2)$.
Both $f(I_R)$ and $g(I_R)$ are concave functions and using the same arguments,
\begin{eqnarray}
F_{\gamma_\textrm{min}}(\bm{\mu})
%\!\!\!\! &=& \!\!\!\! \Pr \left\{ {W\left( {{\gamma _h} + {I_R}} \right) < {\gamma_{_\mathrm{th}}},\frac{{{\gamma _h}}}{{1 + {I_R}}} > {\gamma_{_\mathrm{th}}}}\right\}\nonumber \\
% &&+\Pr \left\{ {\frac{{{\gamma _h}}}{{1 + {I_R}}} < {\gamma_{_\mathrm{th}}}} \right\} \nonumber \\
 &=&\!\!\! {\mathbb{E}_{{I_R}}}\left\{ {1 - \exp \left[ { - \frac{{\gamma \left( {1 + {I_R}} \right)}}{{{{\bar \gamma }_h}}}} \right]} \right\} \nonumber\\
 &&\!+ \mathbb{E}_{W,{I_R}}\Big\{\exp \left[ { - \frac{{\gamma \left( {1 + {I_R}} \right)}}{{{{\bar \gamma }_h}}}} \right]\nonumber\\
  &&- \exp \left[ { - \frac{1}{{{{\bar \gamma }_h}}}\left( {\frac{\gamma }{W} - {I_R}} \right)} \right]  \Big\} \nonumber \\
 &=& {\exp \left( { - \frac{{\rm{1}}}{{{{\bar \gamma }_g}}}} \right)\mathbb{E}_{{I_R}}}\left\{ f(I_R) \right\} +
 \mathbb{E}_{W,{I_R}}\left\{f(I_R)  \right\} \nonumber\\
 &&+ \mathbb{E}_{W,{I_R}}\left\{g(I_R) \right\}
\end{eqnarray}
is also Schur-concave with respect to $\bm{\mu}$.
Since $1 - F_{\gamma_\textrm{min}}(\bm{\mu_1}) \geq  1 - F_{\gamma_\textrm{min}}(\bm{\mu_2})\geq 0$, integration with respect to $\gamma$ gives $C_\mathrm{erg}(\bm{\mu}_1) \geq C_\mathrm{erg}(\bm{\mu}_2)$.
Therefore, the ergodic capacity, $C_\mathrm{erg}$, and accordingly the achievable throughput, $T_\mathrm{erg}$, are Schur-convex with respect to $\bm{\mu}$.
\end{IEEEproof}

According to the Schur-convex property of ergodic capacity,
under different interference power distributions, the corresponding system capacity relationship, i.e., the ordering of capacities, can be obtained. For example, our results imply that the worst scenario for the capacity performance occurs when the received interfering signals are of equal strength at the relay, whereas the best case happens when there is only one interferer affecting the relay.
\section{Outage Capacity }\label{Outage Capacity}
In this section, the exact closed-form expressions of the outage probability, outage capacity and the achievable throughput are derived for the dual-hop energy harvesting DF relaying system. The impact of the interference power distribution on the outage capacity and the achievable throughput is also analyzed, based on the majorization theory.
\subsection{Outage Probability}
As an important performance measure of wireless systems, outage probability is defined as the probability that the instantaneous output SNR falls below a pre-defined threshold $\gamma_{_\mathrm{th}}$.
This SNR threshold guarantees the minimum quality-of-service requirement of the destination users.
Mathematically speaking, $P_{\rm out}(\gamma_{_\mathrm{th}})={\rm Pr}\left\{\gamma \leq \gamma_{_\mathrm{th}}\right\}$.
In the DF relaying system under study, if the received SINR $\gamma_{_{SR}}$ at the relay is below $\gamma_{_\mathrm{th}}$, then the data received over that fading block cannot be decoded correctly with probability approaching $1$, and thus, the receiver at the destination declares an outage since the data will not be transmitted to the destination. Therefore, the outage probability at the destination is composed of two parts, that is,
\newcounter{MYtempeqncnt0}
\begin{figure*}[hb]
\normalsize
\hrulefill
\setcounter{MYtempeqncnt0}{\value{equation}}
\setcounter{equation}{34}
\begin{eqnarray}\label{CDF_D}
P_{\mathrm{out}}\left( \gamma_{_\mathrm{th}} \right)
&=& 1 - \sum\limits_{i = 1}^{\upsilon \left( {\bm {A}} \right)} {\sum\limits_{j = 1}^{{\tau _i}\left( {\bm {A}} \right)} {{\chi _{i,j}}\left( {\bm {A}} \right)} } {\left( {1 - \frac{{{\mu _{\left\langle i \right\rangle }}}}{{{{\bar \gamma }_h}}}} \right)^{ - j}}
\bigg\{\Gamma\Big(1,\frac{\gamma_{_\mathrm{th}}}{\bar \gamma_h};\frac{\gamma_{_\mathrm{th}}}{{\bar \gamma }_h{\bar \gamma }_g}\Big)
-\frac{1}{{\bar \gamma }_h}\exp \left( {\frac{{{a_{\left\langle i \right\rangle }}\gamma_{_\mathrm{th}}}}{{1 + \gamma_{_\mathrm{th}}}}} \right)
\nonumber \\
& & \times
\sum\limits_{k = 0}^{j - 1}\frac{1}{k!}
\left(\frac{{ - a_{\left\langle i \right\rangle} \gamma_{_\mathrm{th}}}}{1 + \gamma_{_\mathrm{th}}}\right)^k
\sum\limits_{m = 0}^k
{k \choose m }
\frac{b_{\left\langle i \right\rangle }^{-1}}{(-b_{\left\langle i \right\rangle }\gamma_{_\mathrm{th}})^m}
\Gamma\Big(m+1,b_{\left\langle i \right\rangle }\gamma_{_\mathrm{th}};\frac{b_{\left\langle i \right\rangle }\gamma_{_\mathrm{th}}}{\bar \gamma_g}\Big)\bigg\}.
\end{eqnarray}
\setcounter{equation}{\value{MYtempeqncnt0}}
\vspace*{0pt}
%\end{figure*}
\newcounter{MYtempeqncnt1}
%\begin{figure*}[hb]
\normalsize
\hrulefill
\setcounter{MYtempeqncnt1}{\value{equation}}
\setcounter{equation}{36}
\begin{eqnarray}\label{Frd2}
P_{\mathrm{out}}\left( \gamma_{_\mathrm{th}} \right)
 &=& 1 - \sum\limits_{i = 1}^{\upsilon \left( {\bm A} \right)} {\sum\limits_{j = 1}^{{\tau _i}\left( {\bm A} \right)} {{\chi _{i,j}}\left( {\bm A} \right)} } {\left( {1 - \frac{{{\mu _{\left\langle i \right\rangle }}}}{{{{\bar \gamma }_h}}}} \right)^{ - j}}
\bigg[
\int_{\gamma_{_\mathrm{th}}}^\infty {\frac{1}{{\bar \gamma }_h}\exp \left( { - \frac{\gamma_{_\mathrm{th}}}{{{{\bar \gamma }_g}z}} - \frac{z}{{{{\bar \gamma }_h}}}} \right)} dz  \nonumber \\
& & - \sum\limits_{k = 0}^{j - 1} {\frac{{a_{\left\langle i \right\rangle }^k}}{{k!}}} \int_{\gamma_{_\mathrm{th}}}^\infty  {\frac{1}{{\bar \gamma }_h} \exp \left( { - \frac{\gamma_{_\mathrm{th}}}{{{{\bar \gamma }_g}z}} - \frac{z}{{{{\bar \gamma }_h}}}} \right)\exp \left( { - {a_{\left\langle i \right\rangle }}\frac{{z - \gamma_{_\mathrm{th}}}}{{1 + \gamma_{_\mathrm{th}}}}} \right){{\left( {\frac{{z - \gamma_{_\mathrm{th}}}}{{1 + \gamma_{_\mathrm{th}}}}} \right)}^k}} dz \bigg].
\end{eqnarray}
\setcounter{equation}{\value{MYtempeqncnt1}}
\vspace*{0pt}
\end{figure*}
\begin{eqnarray} \label{outage1}
P_{\mathrm{out}}\left( {{\gamma_{_\mathrm{th}}}} \right)
 &=& \Pr \left\{ {{\gamma_{_{SR}}} \leq {\gamma_{_\mathrm{th}}}} \right\} + \Pr \left\{ {{\gamma_{_{SR}}} > {\gamma_{_\mathrm{th}}}} \right\}\nonumber\\
 &&\times\Pr \left\{ {\left. {{\gamma_{_{RD}}} \leq {\gamma_{_\mathrm{th}}}} \right|{\gamma_{_{SR}}} > {\gamma_{_\mathrm{th}}}} \right\} \nonumber\\
 &=& \Pr \left\{ {W\left( {{\gamma _h} + {I_R}} \right) \leq {\gamma_{_\mathrm{th}}},\frac{{{\gamma _h}}}{{1 + {I_R}}} > {\gamma_{_\mathrm{th}}}}\right\}  \nonumber \\
 &&+
 \Pr \left\{ {\frac{{{\gamma _h}}}{{1 + {I_R}}} \leq {\gamma_{_\mathrm{th}}}} \right\}\label{outage1b} \\
 &=& \Pr \big \{ W(\gamma_{h} + I_R)\mathbbm{1}_{\mathcal{C}} \leq {\gamma_{_\mathrm{th}}}\big \}\label{outage1c},
\end{eqnarray}
where $\mathbbm{1}_{\mathcal{C}}$ is the indicator random variable for the set
$\mathcal{C}=\{\gamma_{_{SR}} > \gamma_{_\mathrm{th}}\}$, i.e.,
$\mathbbm{1}_{\mathcal{C}}=1$ if $\gamma_{_{SR}} > \gamma_{_\mathrm{th}}$, otherwise, $\mathbbm{1}_{\mathcal{C}}=0$.

Note that, in contrast to traditional DF relaying system with no rechargeable nodes, the transmission power $P_{_R}$ at the relay in the energy harvesting system is a random variable, which depends on the replenished energy from the interference and information signal.
Therefore, the distribution of the received SNR at the destination is determined not only by the distribution of the relay-to-destination channel power gain $|g|^2$, but also by the distribution of the information and interference signal power, i.e., $\gamma_{h}$ and $I_R$. On the other hand, in the counterpart system of conventional DF relaying, when the relay can decode the information correctly, its transmission power $P_{_R}$ is a constant, and thus, the received SNR at the destination only depends on the relay-to-destination channel power gain $|g|^2$.

\vspace*{-0.in}\begin{mytheorem}\label{PrPower}
 Define $Z \triangleq (\gamma_{h} + I_R)\mathbbm{1}_{\mathcal{C}}$, then the PDF of $Z$ is given by
\begin{eqnarray}\label{PDFZ}
f_Z(z)
\!\!\!\!&=&\!\!\!\!\mathbbm{1}_{\mathcal{Z}}
\frac{1}{{{{\bar \gamma }_h}}}\exp ( - \frac{z}{{{{\bar \gamma }_h}}})\sum\limits_{i = 1}^{\upsilon \left( {\bm {A}} \right)} {\sum\limits_{j = 1}^{{\tau _i}\left( {\bm {A}} \right)} {{\chi _{i,j}}\left( {\bm {A}} \right)} } {\left( {{\rm{1}} - \frac{{{\mu _{\left\langle i \right\rangle }}}}{{{{\bar \gamma }_h}}}} \right)^{ - j}}\nonumber\\
\!\!\!&\times&\!\!\!\! \!\!\!
\left[{{\rm{1}}\!-\! \exp \left( {\!\! - {a_{\left\langle i \right\rangle }}\frac{{z - \gamma_{_\mathrm{th}}}}{{1 + \gamma_{_\mathrm{th}}}}} \right)\sum\limits_{k = 0}^{j - 1} {\frac{{a_{\left\langle i \right\rangle }^k}}{{k!}}} {{\left( {\frac{{z - \gamma_{_\mathrm{th}}}}{{1 + \gamma_{_\mathrm{th}}}}} \right)}^k}} \right]
\label{fZ3}
\end{eqnarray}
where $a_{\left\langle i \right\rangle }\triangleq \frac{{\rm{1}}}{{{\mu _{\left\langle i \right\rangle }}}} - \frac{{\rm{1}}}{{{\bar \gamma }_h}}$ and $\mathbbm{1}_{\mathcal{Z}}$ is the indicator random variable for the set
$\mathcal{Z}=\{z> \gamma_{_\mathrm{th}}\}$, i.e.,
$\mathbbm{1}_{\mathcal{Z}}=1$ if $ z > \gamma_{_\mathrm{th}}$, otherwise, $\mathbbm{1}_{\mathcal{Z}}=0$.
\end{mytheorem}
\begin{IEEEproof}
The CDF of the random variable $Z = (\gamma_{h} + I_R)  \mathbbm{1}_{\mathcal{C}}$ is given by
\begin{eqnarray}\label{CDFZ}
F_Z(z)
&=& \Pr\{Z\leq z\} \nonumber \\
&=& \int\int_{{x,y}\in \mathcal{S}}f_{\gamma_{h},I_R}(x,y) \mathrm{d}x \mathrm{d}y,
\end{eqnarray}
where the set $\mathcal{S}={\left\{ {x + y \leq z,{\kern 1pt} \frac{x}{{1 + y}} > \gamma_{_\mathrm{th}}}, x\geq0, y\geq0 \right\}}$.
After some set manipulations, we have
$\mathcal{S}\neq \emptyset $ if and only if $ z> \gamma_{_\mathrm{th}}$.
Since $\gamma_{h}$ and $I_R$ are independent, we get the joint distribution $f_{\gamma_{h},I_R}(x,y) = f_{\gamma_{h}}(x)f_{I_R}(y)$.
Then, after some straightforward algebraic derivations, we obtain
\begin{equation}\label{Fz}
F_Z(z) =
\mathbbm{1}_{\mathcal{Z}}
\int_0^{ \frac{z - \gamma_{_\mathrm{th}}}{1 + \gamma_{_\mathrm{th}}}}
\int_{(1+y)\gamma_{_\mathrm{th}}}^{z - y} {f_{\gamma_{h}}(x)f_{I_R}(y)} \mathrm{d}x \mathrm{d}y.
\end{equation}
Now, substituting (\ref{pdfX}) and (\ref{pdfY}) into (\ref{Fz}) and integrating with respect to $x$ and $y$ yields the CDF of $Z$. Then the PDF of $Z$ follows directly from differentiating $F_Z(z)$ with respect to $z$. Here \cite[Eq.(3.351.1)]{Gradshteyn} was used to reach (\ref{PDFZ}).
\end{IEEEproof}

Next, we evaluate the outage probability at the destination by using the above Theorem \ref{PrPower}.
\begin{mytheorem}\label{D-SNR}
The outage probability at the destination node of the interference aided energy harvesting DF relaying system is given by (\ref{CDF_D}) shown at the bottom of the page,
where
${b_{\left\langle i \right\rangle }} = \frac{1}{{{{\bar \gamma }_h}}} + \frac{{{a_{\left\langle i \right\rangle }}}}{{1 + \gamma_{_\mathrm{th}}}}$ and $\Gamma(a,x;b)$ is the generalized incomplete Gamma function defined by
$\Gamma(a,x;b) \triangleq \int_{x}^{\infty}t^{a-1}\exp(-t-bt^{-1})d t$.
\end{mytheorem}
\begin{IEEEproof}
We have
\setcounter{equation}{35}
\begin{eqnarray}\label{Frd}
P_{\mathrm{out}}\left( \gamma_{_\mathrm{th}} \right)
 &=& \Pr \{WZ \leq \gamma_{_\mathrm{th}}\} \nonumber \\
 &=& \mathbb{E}_Z \left\{ {1 - \exp \left( - \frac{\gamma_{_\mathrm{th}}}{{{{\bar \gamma }_g}Z}} \right)} \right\} \nonumber \\
 &=& 1 - \int_0^\infty  {\exp \left( { - \frac{\gamma_{_\mathrm{th}}}{{{{\bar \gamma }_g}z}}} \right){f_Z}} \left( z \right)dz.
 \end{eqnarray}
According to Theorem \ref{PrPower}, by substituting (\ref{PDFZ}) into (\ref{Frd}), we obtain (\ref{Frd2}) shown at the bottom of the page.
Next, we focus on the two integrations in (\ref{Frd2}).
For the first integration, from the definition of the generalized incomplete Gamma function, we have
\setcounter{equation}{37}
\begin{equation} \label{A1}
\int_{\gamma_{_\mathrm{th}}}^\infty {\frac{1}{{\bar \gamma }_h}\exp \left( { - \frac{\gamma_{_\mathrm{th}}}{{{{\bar \gamma }_g}z}} - \frac{z}{{{{\bar \gamma }_h}}}} \right)} dz
=\Gamma \Big(1,\frac{\gamma_{_\mathrm{th}}}{\bar \gamma_h};\frac{\gamma_{_\mathrm{th}}}{{\bar \gamma }_h{\bar \gamma }_g}\Big).
\end{equation}
For the second integration, exploiting the Taylor series expansion of
$(z-\gamma_k)^k$ with respect to $z$ and the identity of the generalized incomplete Gamma function lead to (\ref{CDF_D}).
\end{IEEEproof}

Simplified expressions for the outage probability as 1) the interferences are i.i.d. or 2) the number of interferers equals one, are derived and given by
\begin{eqnarray}\label{CDF_D1}
P_{\mathrm{out}}^{(1)}\left( \gamma_{_\mathrm{th}} \right)
\!\!\!\!&=&\!\!\!\! 1 - {\left( {1 - \frac{{{\mu}}}{{{{\bar \gamma }_h}}}} \right)^{ - M}}
\bigg\{\Gamma\Big(1,\frac{\gamma_{_\mathrm{th}}}{\bar \gamma_h};\frac{\gamma_{_\mathrm{th}}}{{\bar \gamma }_h{\bar \gamma }_g}\Big)
-\frac{1}{{\bar \gamma }_h}
\nonumber \\
\!\!\!\!& &\!\!\!\! \times\exp \left( {\frac{{{a}\gamma_{_\mathrm{th}}}}{{1 + \gamma_{_\mathrm{th}}}}} \right)
\sum\limits_{k = 0}^{M - 1}\frac{1}{k!}
\left(\frac{{ - a \gamma_{_\mathrm{th}}}}{1 + \gamma_{_\mathrm{th}}}\right)^k
\sum\limits_{m = 0}^k \nonumber \\
\!\!\!\!& &\!\!\!\! {k \choose m }
\frac{b^{-1}}{(-b\gamma_{_\mathrm{th}})^m}
\Gamma\Big(m+1,b\gamma_{_\mathrm{th}};\frac{b\gamma_{_\mathrm{th}}}{\bar \gamma_g}\Big)\bigg\}
\end{eqnarray}
and
\begin{eqnarray}\label{CDF_N1}
P_{\mathrm{out}}^{(2)}\left( \gamma_{_\mathrm{th}} \right)
\!\!\!\!&=&\!\!\!\! 1 - \frac{{\bar \gamma }_h}{{\bar \gamma }_h-\mu}
\Gamma\Big(1,\frac{\gamma_{_\mathrm{th}}}{\bar \gamma_h};\frac{\gamma_{_\mathrm{th}}}{{\bar \gamma }_h{\bar \gamma }_g}\Big)
+\frac{b^{-1}}{{\bar \gamma }_h-\mu}
\nonumber \\
\!\!\!\!& &\!\!\!\! \times
\exp \left( {\frac{{{a}\gamma_{_\mathrm{th}}}}{{1 + \gamma_{_\mathrm{th}}}}} \right)
\Gamma\Big(1,b\gamma_{_\mathrm{th}};\frac{b\gamma_{_\mathrm{th}}}{\bar \gamma_g}\Big)
\end{eqnarray}
respectively,
where $a\triangleq \frac{{\rm{1}}}{{{\mu}}} - \frac{{\rm{1}}}{{{\bar \gamma }_h}}$ and ${b} \triangleq \frac{1}{{{{\bar \gamma }_h}}} + \frac{{{a}}}{{1 + \gamma_{_\mathrm{th}}}}$.
\subsection{Outage Capacity and Achievable Throughput}
Outage capacity, in the unit of bit/s/Hz, is defined as the maximum constant rate that can be maintained over fading blocks with a specified outage probability. It is used for slowly varying channels, where the instantaneous SNR $\gamma$ is assumed to be constant for a large number of symbols.
In the DF-cooperative communication system under study, the outage capacity in the unit of bit/s/Hz is expressed as
\begin{equation} \label{Outage Capacity}
C_\mathrm{out}= \frac{1}{2}\left[1-P_{\mathrm{out}}( \gamma_{_\mathrm{th}} )\right]\log_2(1+\gamma_{_\mathrm{th}}).
\end{equation}
The factor $1/2$  accounts for the fact that two transmission phases are involved in the communication between the source $S$ and the destination $D$.
The achievable throughput at the destination relates only to the effective information transmission time and is then given by
\begin{equation}\label{through_TS}
T_\mathrm{out}
 = (1-\alpha)C_\mathrm{out},
\end{equation}
for the system employing time switching, and by
\begin{equation}\label{through_PS}
T_\mathrm{out} = C_\mathrm{out},
\end{equation}
for the system implementing power splitting.
\subsection{Impact of Interference Power Distribution}
\begin{mytheorem}\label{Concave_Out}
The outage capacity, $C_\mathrm{out}$, and the achievable throughput, $T_\mathrm{out}$, of the interference aided energy-harvesting DF relaying system is Schur-convex with respect to $\bm{\mu}$, where $\bm{\mu}=(\mu _{1},\mu _{2},\ldots ,\mu _{M})$ with $\mu_{1}\geq \mu_{2}\geq \cdots \geq \mu_{M}\geq 0$.
\end{mytheorem}
\begin{IEEEproof}
According to (\ref{outage1b}) and using the same arguments as in the proof of Theorem \ref{Concave_Erg}, we can see that the outage capacity $C_\mathrm{out}$ and accordingly the achievable throughput $T_\mathrm{out}$ are also Schur-convex with respect to $\bm{\mu}$.
\end{IEEEproof}

Note that Theorem 1 and Theorem 4 provide engineering insights for design of energy harvesting relay system.
With regard to application, the proposed energy harvesting relay system can be seen as building block of a larger cellular network.
For instance, consider full frequency reuse for all base stations, which is studied extensively recently for multi-cell cooperation, and where a base station serves farther users through the help of intermediate relaying nodes. From a system design point-of-view, how to choose the relay location to obtain the largest capacity is a meaningful and challenging problem.
Relays that are positioned at different geometric locations may suffer the same total received interference power (at the same contour) from neighboring base stations, but with different power distributions.
Based on the analysis provided in this paper, the best relay positioning can be identified. Definitely, the detailed application depends on the specific problem, which is beyond the scope of the paper, and can be considered in future extensions of this work.
\section{Numerical Results and Discussions}\label{simulation}
In this section, numerical examples are presented and corroborated by simulation results to examine the throughput, $T_\mathrm{erg}$ and $T_\mathrm{out}$, of the DF cooperative communication system, where the energy-constrained relay harvests energy from the received information signal and the CCI signals.
Hereafter, and unless stated otherwise, the number of CCI signals at the relay, $M$, is set to $2$ with normalized $\bm{\hat{\mu}}=\frac{\bm{\mu}}{\mathbb{E}\{ {I}_R \}}=(0.6, 0.4)$. The threshold $\gamma_{_\mathrm{th}}$ is set to $8\mathrm{dB}$ and the energy conversion efficiency $\eta$ is set to $1$. To better evaluate the effects of the interferences on the system's throughput, we define $\frac{P_{_S}\Omega_h}{\sum_{i = 1}^M P_i\Omega_{\beta_i}}$ as the average signal-to-interference ratio (SIR) at the relaying node and $\frac{P_{_S}\Omega_h}{\sigma_R^2}$ as the first-hop average SNR.

\begin{figure}[!ht]
  \centering
\vspace*{-0.2in}\epsfig{figure=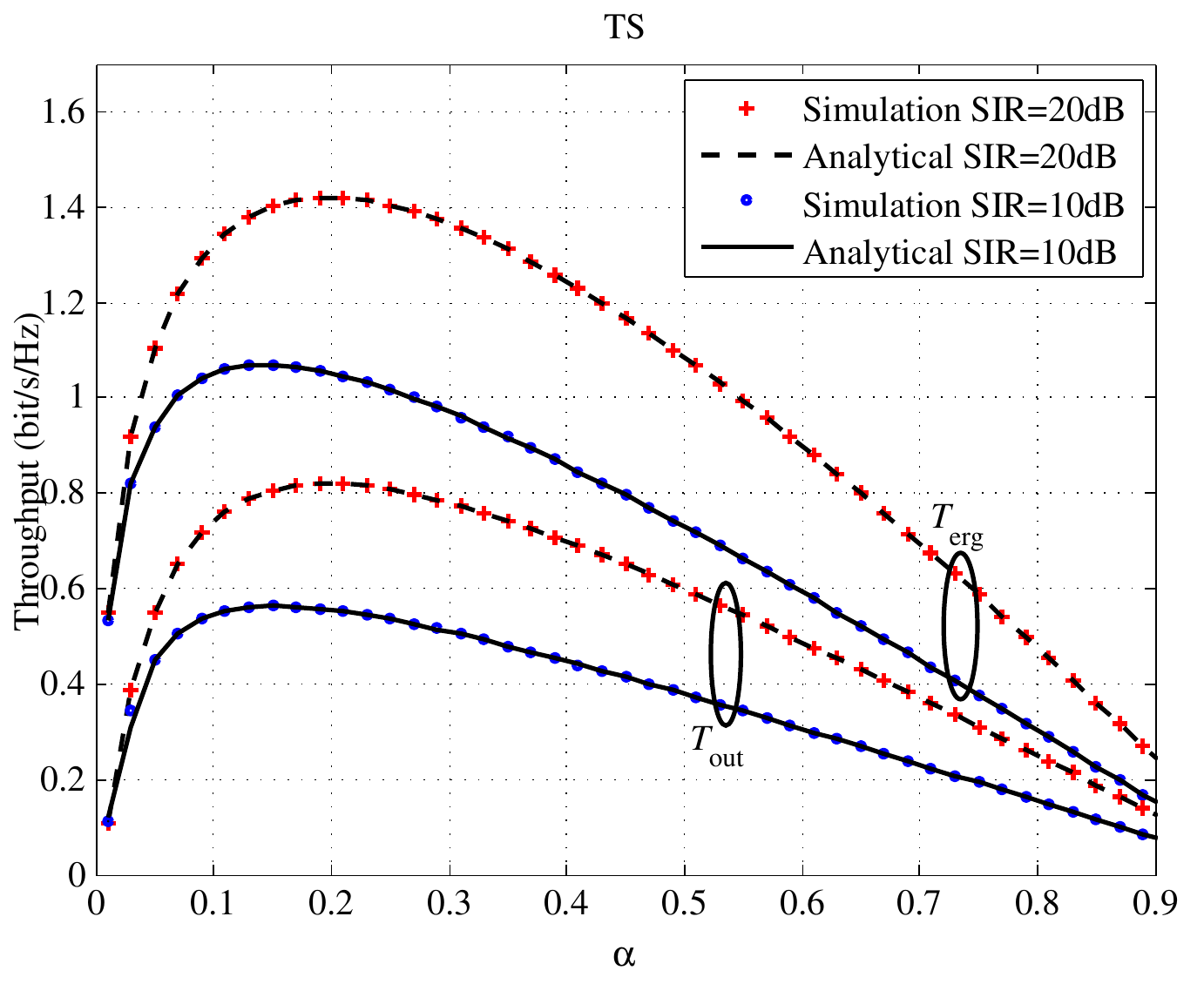,width=3.1in}\\
\vspace*{-0.1in}\caption{Throughput $T_\mathrm{erg}$ and $T_\mathrm{out}$ versus the energy harvesting ratio $\alpha$ for different values of average SIR received at the relay, where the first-hop average SNR is $20\mathrm{dB}$.  }
\label{Fig2}
\end{figure}
\begin{figure}[!ht]
  \centering
\vspace*{-0.0in}\epsfig{figure=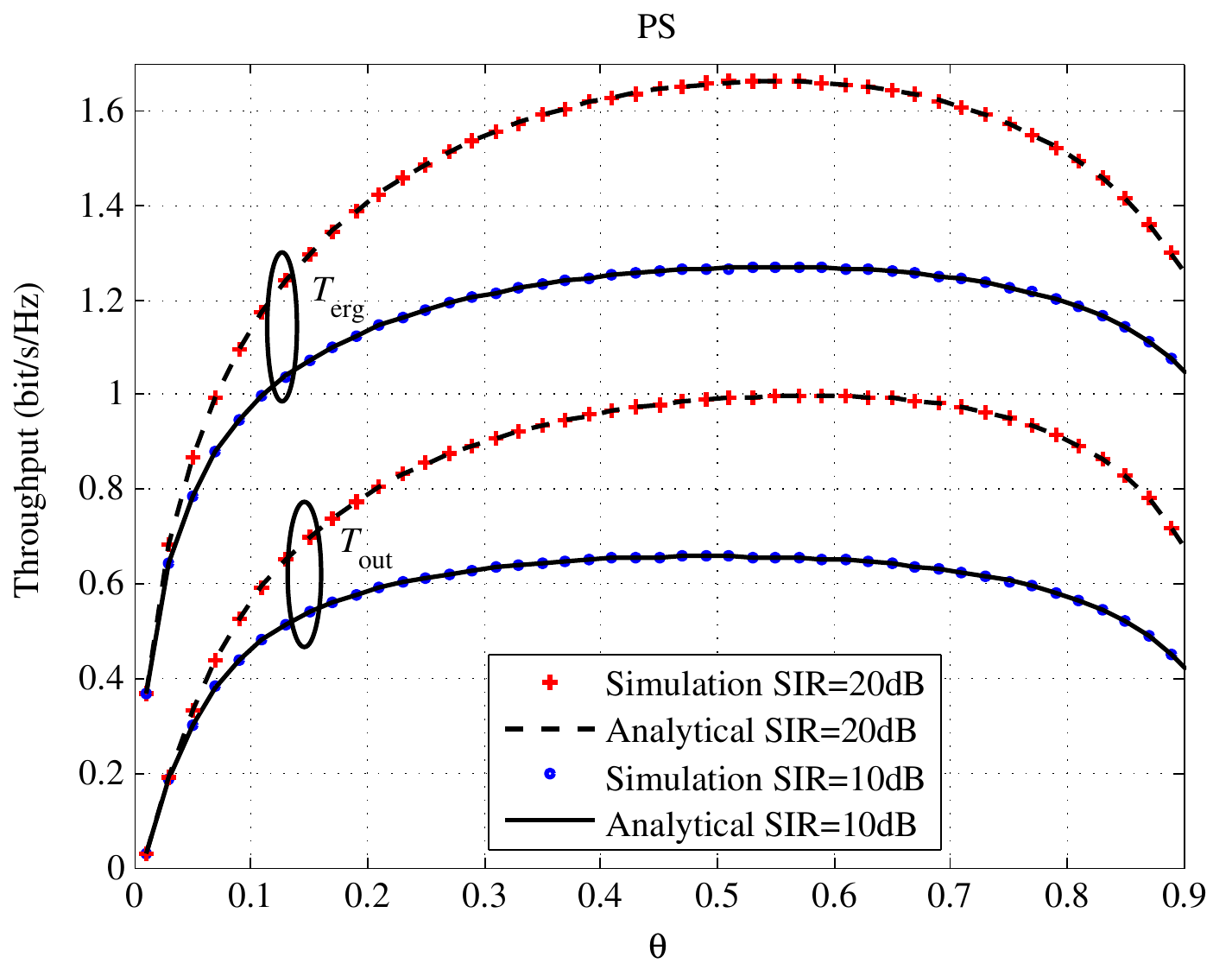,width=3.1in}\\
\vspace*{-0.1in}\caption{Throughput $T_\mathrm{erg}$ and $T_\mathrm{out}$ versus the energy harvesting ratio $\theta$ for different values of average SIR received at the relay, where the first-hop average SNR is $20\mathrm{dB}$. }
\label{Fig3}
\end{figure}

For the system with TS protocol, Fig. \ref{Fig2} shows the throughput $T_\mathrm{erg}$ and $T_\mathrm{out}$ versus the energy harvesting ratio $\alpha$ for different values of average SIR received at the relay, where the first-hop average SNR is $20\mathrm{dB}$.
It is observed that the analytical results of (\ref{throughput_TS}) and (\ref{through_TS})  match well the simulation results.
As the energy harvesting ratio $\alpha$ increases from $0$ to $1$, the throughput of the system increases at first until $\alpha$ reaches the optimal value where the throughput gets its maximum, and thereafter decreases from the maximum to zero.
The concave feature of the curves is due to the fact that the energy harvested for the second-hop transmission increases with increasing $\alpha$, which effectively decreases the outage and enhances the capacity of the second hop and, accordingly, improves the throughput of the system.
Meanwhile, as $\alpha$ increases, more data are wasted on energy harvesting and less information is decoded for information transmission which heavily reduces the throughput of the system, therefore, the throughput reaches a maximum and then drops down.
As SIR increases, the optimal throughput and the optimal $\alpha$ both increase. This means that when the received average SNR at the relay is fixed, an increase in the power of the CCI signals can deteriorate the system performance, but effectively reduces the optimal $\alpha$ required to achieve the optimal throughput.

For comparison purposes, Fig. \ref{Fig3} depicts the throughput $T_\mathrm{erg}$ and $T_\mathrm{out}$ versus the energy harvesting ratio $\theta$ for the system with PS protocol under the same simulation settings.
It is observed that the analytical results of (\ref{throughput_PS}) and (\ref{through_PS}) match perfectly the simulation results.
The concave feature of the curves is due to the fact that the energy harvested for the second-hop transmission increases with increasing $\theta$, which effectively decreases the outage and enhances the capacity of the second hop and, accordingly, improves the throughput of the system.
Meanwhile, as $\theta$ increases, more power is harvested for information transmission and less power is left for information decoding which deteriorates the throughput of the system and, thus, the throughput reaches a maximum and then drops down.
In both plots (Fig. 2 and Fig. 3), it is seen that the throughput $T_\mathrm{out}$ is less than the throughput $T_\mathrm{erg}$ due to the requirement of the outage capacity that a fixed date rate is maintained in all non-outage channel states.

\begin{figure}[!ht]
  \centering
\vspace*{-0.2in}\epsfig{figure=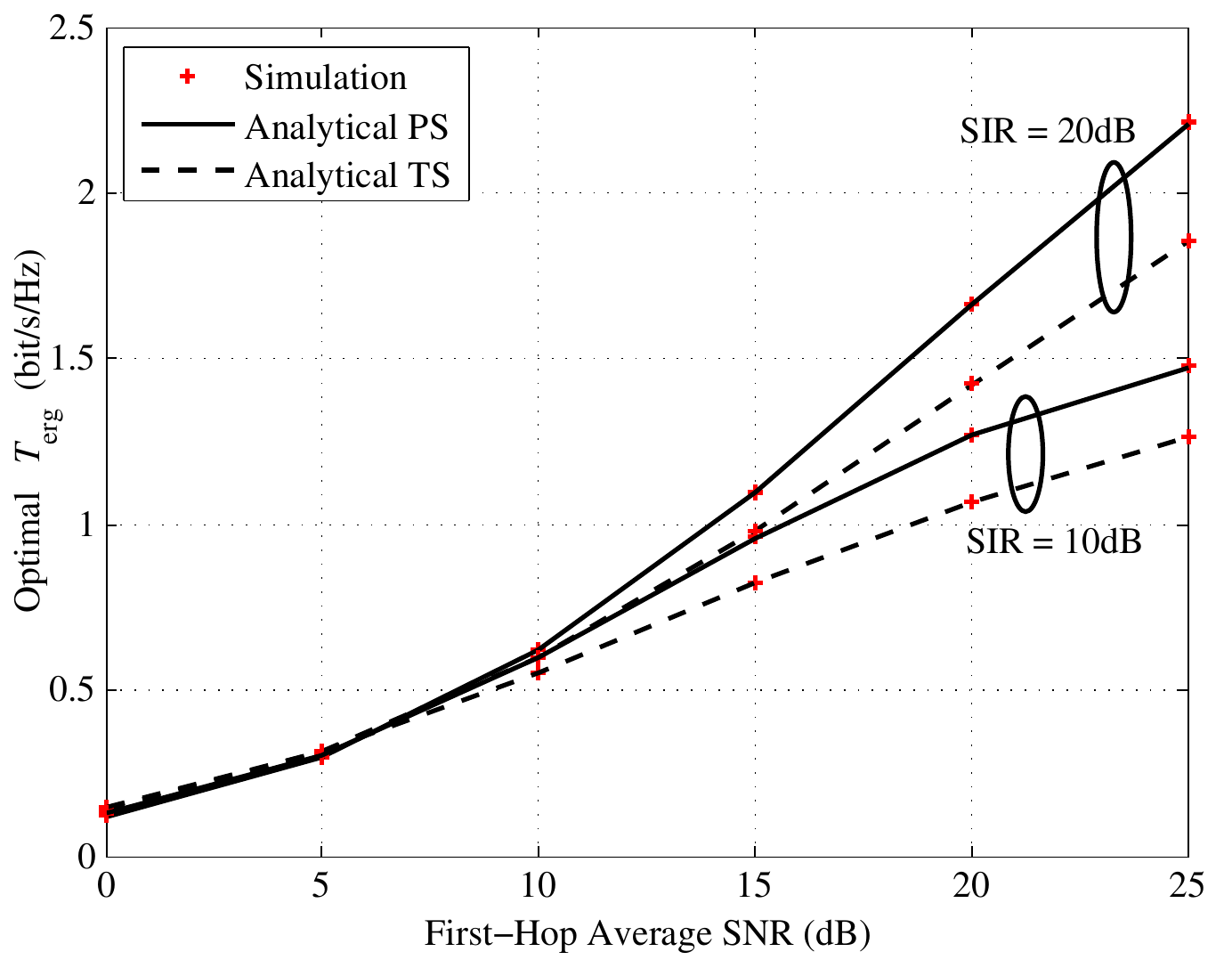,width=3.1in}\\
\vspace*{-0.1in}\caption{Optimal throughput $T_\mathrm{erg}$ versus the first-hop average SNR at different values of average SIR, for both power-splitting (PS) and time-switching (TS).}
\label{Fig4}
\end{figure}

\begin{figure}[!ht]
  \centering
\vspace*{-0.2in}\epsfig{figure=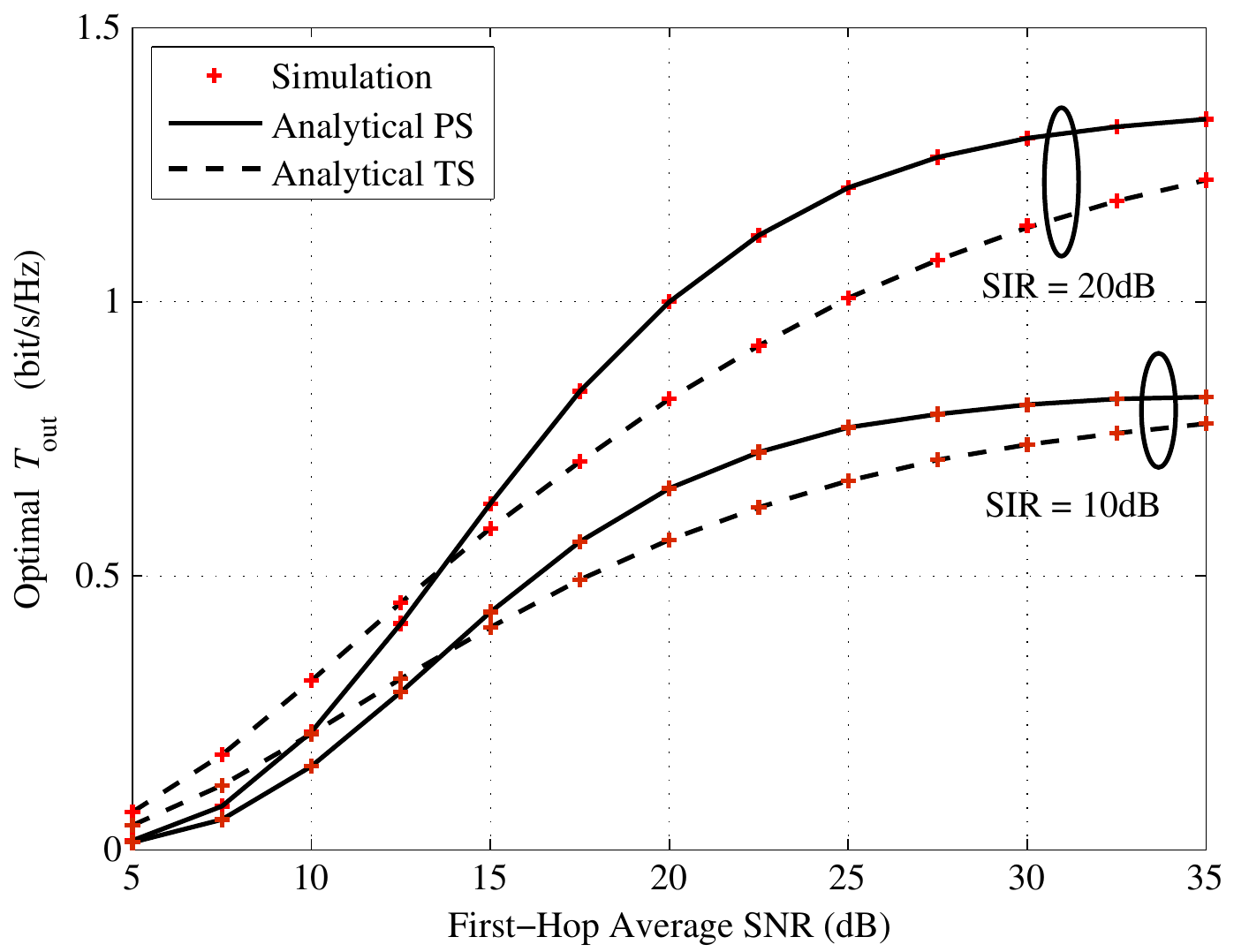,width=3.1in}\\
\vspace*{-0.1in}\caption{Optimal throughput $T_\mathrm{out}$ versus the first-hop average SNR at different values of average SIR, for both PS and TS.}
\label{Fig5}
\end{figure}

Next, we compare the throughput performances of the energy harvesting systems with TS and PS protocols, respectively, to facilitate the choice of these two schemes for designing energy harvesting system. Figures \ref{Fig4} and \ref{Fig5} illustrate the optimal $T_\mathrm{erg}$ and the optimal $T_\mathrm{out}$ versus the first-hop average SNR, respectively, for these two protocols given different values for the average SIR at the relay.
It is observed that the PS protocol is superior to the TS protocol at high SNR, in terms of optimal $T_\mathrm{erg}$ and $T_\mathrm{out}$. At relatively low SNR, on the other hand, the TS-based scheme outperforms the PS one in terms of optimal $T_\mathrm{out}$, but with little difference in optimal $T_\mathrm{erg}$.
This can be explained as follows. At high SNR, power-splitting with optimal ratio $\theta$ ($\theta$ is around half) would not decrease the received SNR significantly so that the information could still be correctly decoded at the relaying node, but for the time-switching scheme, there always exists an information loss at the energy harvesting phase.
Similarly, at low SNR, power splitting with optimal ratio $\theta$ would lead to more decoding errors at the relaying node.

\begin{figure}[!ht]
  \centering
\vspace*{-0.2in}\epsfig{figure=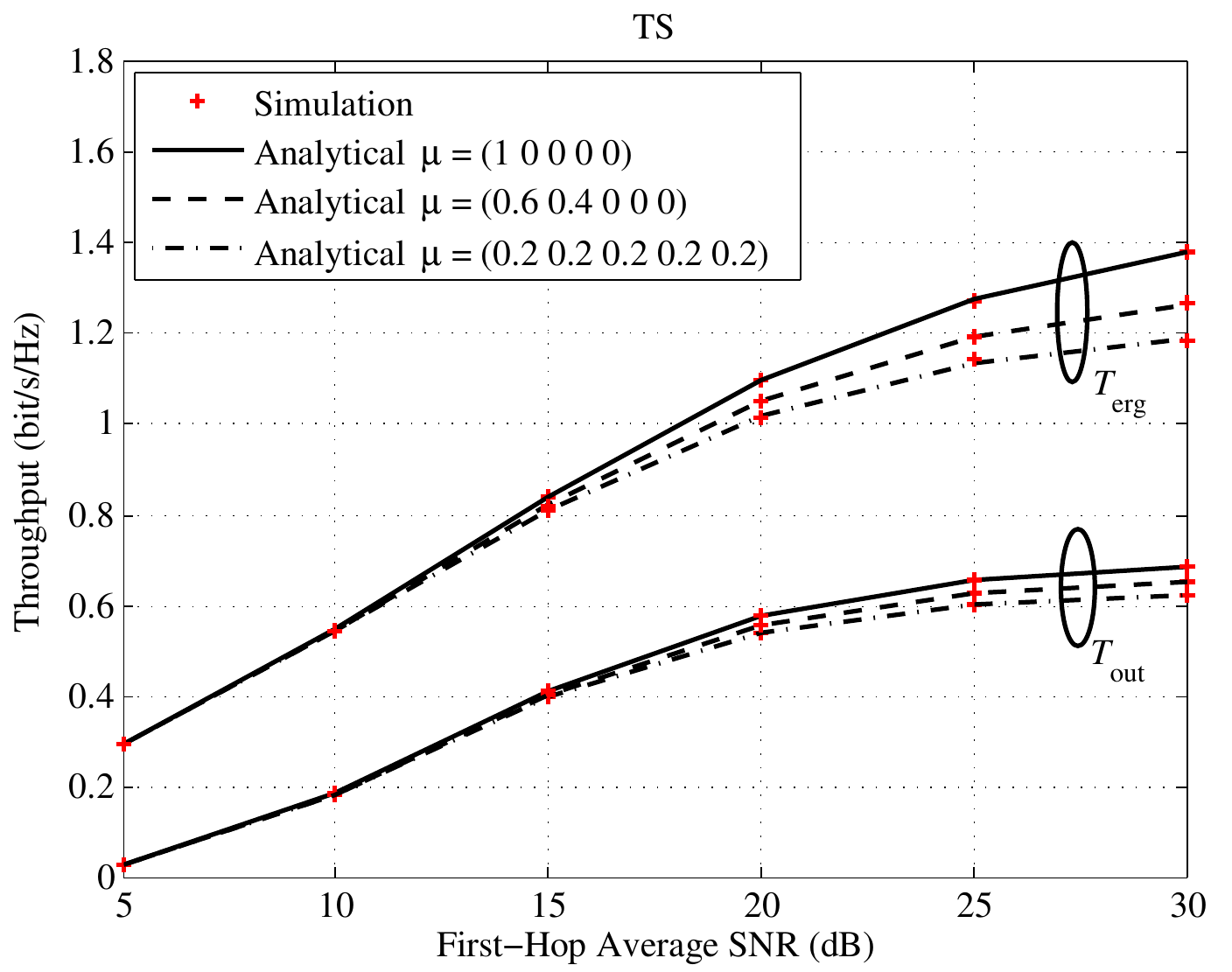,width=3.1in}\\
\vspace*{-0.1in}\caption{Throughput of the TS based system, $T_\mathrm{erg}$ and $T_\mathrm{out}$, versus the first-hop average SNR under different CCI power distributions.}
\label{Fig6}
\end{figure}
\begin{figure}[!ht]
  \centering
\vspace*{-0.0in}\epsfig{figure=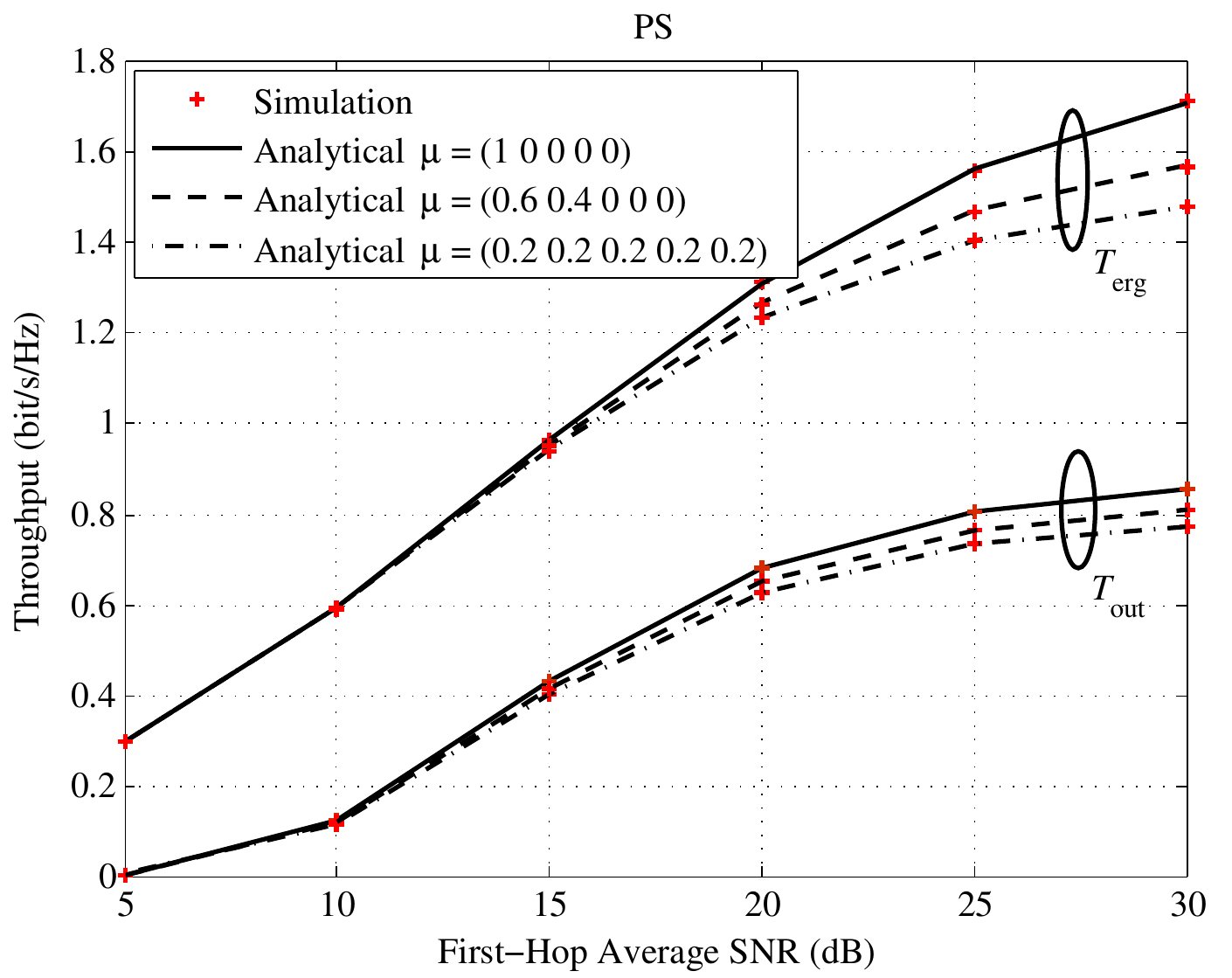,width=3.1in}\\
\vspace*{-0.1in}\caption{Throughput of the PS based system, $T_\mathrm{erg}$ and $T_\mathrm{out}$, versus the first-hop average SNR under different CCI power distributions.}
\label{Fig7}
\end{figure}

The impact of the interference power distribution on the throughput performance is shown in Fig. \ref{Fig6} for the system with TS protocol and in Fig. \ref{Fig7} for the system with PS protocol. The energy harvesting ratios $\alpha$ and $\theta$ are set to $0.2$ and $0.6$, respectively. The SIR at the relay is $10\mathrm{dB}$. The total interference power is the same at each SNR but with different normalized power distribution: $\bm{\hat{\mu}}_{1}=(1,0,0,0,0)$,
$\bm{\hat{\mu}}_{2}=(0.6,0.4,0,0,0)$ and $\bm{\hat{\mu}}_{3}=(0.2,0.2,0.2,0.2,0.2)$. According to the definition of majorization, we have $\bm{\mu}_{1}\succeq \bm{\mu}_{2}\succeq \bm{\mu}_{3}$ and thus, $T_\mathrm{erg}(\bm{\mu}_{1})\geq T_\mathrm{erg}(\bm{\mu}_{2})\geq T_\mathrm{erg}(\bm{\mu}_{3})$ and $T_\mathrm{out}(\bm{\mu}_{1})\geq T_\mathrm{out}(\bm{\mu}_{2})\geq T_\mathrm{out}(\bm{\mu}_{3})$, since the throughput is Schur-convex with respect to $\bm{\mu}$ as proven by Theorems \ref{Concave_Erg} and \ref{Concave_Out}. This is clearly shown by the simulation results, and it implies that the worst scenario for the throughput performance occurs when the interfering signals are of equal received power at the relay, whereas the best case happens when there is only one interferer affecting the relay.

\section{Conclusion}\label{conclusions}

In this paper, an RF-based energy harvesting relaying system was proposed, where the energy-constrained relay harvests energy from the superposition of received information signal and co-channel interference (CCI)  signals, and then uses that harvested energy to forward the correctly decoded signal to the destination.
The time-switching (TS) and the power-splitting (PS) protocols were adopted, and their ensuing performance was compared.
Different from traditional decode-and-forward relaying system with no rechargeable nodes, the transmission power of the energy constrained relay is not a constant anymore but a random variable depending on the variation of available energy harvested from the received information and CCI signals at the relay.
Analytical expressions for the ergodic capacity as well as for the outage capacity were derived to determine the corresponding system achievable throughputs.
The PS scheme was demonstrated to be superior to TS at high SNR in terms of the achievable throughput from ergodic or outage capacity, while at relatively low SNR, TS outperforms PS in terms of the achievable throughput from outage capacity.
Furthermore, considering
different interference power distributions with equal aggregate interference power at the relay, the corresponding system capacity relationship,
i.e., the ordering of capacities, was obtained. The results reveal that the worst scenario for the capacity performance occurs when the received interfering signals are of equal strength
at the relay, whereas the best case occurs when there is only one interferer affecting the relay.

\vfill

\flushend


\begin{thebibliography}{10}
\providecommand{\url}[1]{#1}
\csname url@samestyle\endcsname
\providecommand{\newblock}{\relax}
\providecommand{\bibinfo}[2]{#2}
\providecommand{\BIBentrySTDinterwordspacing}{\spaceskip=0pt\relax}
\providecommand{\BIBentryALTinterwordstretchfactor}{4}
\providecommand{\BIBentryALTinterwordspacing}{\spaceskip=\fontdimen2\font plus
\BIBentryALTinterwordstretchfactor\fontdimen3\font minus
  \fontdimen4\font\relax}
\providecommand{\BIBforeignlanguage}[2]{{%
\expandafter\ifx\csname l@#1\endcsname\relax
\typeout{** WARNING: IEEEtran.bst: No hyphenation pattern has been}%
\typeout{** loaded for the language `#1'. Using the pattern for}%
\typeout{** the default language instead.}%
\else
\language=\csname l@#1\endcsname
\fi
#2}}
\providecommand{\BIBdecl}{\relax}
\BIBdecl

\bibitem{WSN}
S.~Misra, M.~Reisslein, and G.~Xue, ``A survey of multimedia streaming in
  wireless sensor networks,'' \emph{{IEEE} Commun. Surveys Tuts.}, vol.~10,
  no.~4, pp. 18--39, 2008.

\bibitem{Monitor}
F.~Hu, Y.~Xiao, and Q.~Hao, ``Congestion-aware, loss-resilient bio-monitoring
  sensor networking for mobile health applications,'' \emph{{IEEE} J. Sel.
  Areas Commun.}, vol.~27, no.~4, pp. 450--465, May 2009.

\bibitem{Proceedings}
J. Du, X. Liu, and L. Rao ``Dedicated short-range communications ({DSRC}) standards in the
  united states,'' \emph{{IEEE/ACM} Transactions on Networking}, vol.~99, no.~99, pp. 1--12,  2018.
  
\bibitem{Green}
Z.~Hasan, H.~Boostanimehr, and V.~Bhargava, ``Green cellular networks: A
  survey, some research issues and challenges,'' \emph{{IEEE} Commun. Surveys
  Tuts.}, vol.~13, no.~4, pp. 524--540, 2011.

\bibitem{HuangTWC}
K.~Huang and V.~Lau, ``Enabling wireless power transfer in cellular networks:
  Architecture, modeling and deployment,'' \emph{{IEEE} Trans. Wireless
  Commun.}, vol.~13, no.~2, pp. 902--912, Feb. 2014.

\bibitem{XiaICC}
J.~Du, S. Ma, Y.-C. Wu, and V.~Poor, ``Distributed hybrid power state estimation under PMU sampling phase errors,'' in \emph{IEEE Transactions on Signal Processing}, vol.~62, no.~16, pp. 4052--4063, 2014.
 
\bibitem{XiaICC2}
J.~Du, S. Ma, Y.-C. Wu, and V.~Poor, ``Distributed Bayesian hybrid power state estimation with PMU synchronization errors,'' in \emph{Global Communications Conference (GLOBECOM), 2014 IEEE},  pp. 3174--3179, 2014.




\bibitem{XiaTSP}
M.~{X}ia and S.~Aissa, ``{O}n the efficiency of far-field wireless power
  transfer,'' \emph{IEEE Trans. Signal Processing}, vol.~63, no.~11, pp.
  2835--2847, June 2015.

\bibitem{EH1}
L.~Mateu and F.~Moll, ``Review of energy harvesting techniques and applications
  for microelectronics,'' in \emph{Proc. SPIE Circuits and Syst. II}, 2005, pp.
  359--373.

\bibitem{EH5}
S.~Sudevalayam and P.~Kulkarni, ``Energy harvesting sensor nodes: Survey and
  implications,'' \emph{{IEEE} Commun. Surveys Tuts.}, vol.~13, no.~3, pp.
  443--461, 2011.

\bibitem{wangpower}
Q.~Wang, X.~Liu, J. Du, and F. Kong ``Smart charging for electric vehicles: A survey from the algorithmic perspective,'' \emph{{IEEE} Commun. Surveys Tuts.}, vol.~18, no.~2, pp.
1500--1517, 2016.


\bibitem{EH2}
V.~Raghunathan, S.~Ganeriwal, and M.~Srivastava, ``Emerging techniques for long
  lived wireless sensor networks,'' \emph{{IEEE} Commun. Mag.}, vol.~44, no.~4,
  pp. 108--114, Apr. 2006.

\bibitem{Harvest-Use2}
R.~Ramachandran, V.~Sharma, and P.~Viswanath, ``Capacity of gaussian channels
  with energy harvesting and processing cost,'' \emph{{IEEE} Trans. Inf.
  Theory}, vol.~60, no.~5, pp. 2563--2575, May 2014.

\bibitem{SIGCOMMbestpaper}
V.~Liu, A.~Parks, V.~Talla, S.~Gollakota, D.~Wetherall, and J.~R. Smith,
  ``Ambient backscatter: Wireless communication out of thin air,'' in
  \emph{Proc. ACM SIGCOMM}, Aug. 2013, pp. 39--50.

\bibitem{EH3}
T.~Paing, J.~Shin, R.~Zane, and Z.~Popovic, ``Resistor emulation approach to
  low-power {RF} energy harvesting,'' \emph{{IEEE} Trans. Power Electron.},
  vol.~23, no.~3, pp. 1494--1501, May 2008.

\bibitem{EH4}
R.~Rajesh, V.~Sharma, and P.~Viswanath, ``Information capacity of energy
  harvesting sensor nodes,'' in \emph{Proc. IEEE Int. Symp. Inf. Theory}, July
  2011, pp. 2363--2367.

\bibitem{Ada}
J.~S. Hoa, A.~J. Yeha, E.~Neofytoub, S.~Kima, T.~Tanabea, B.~Patlollab,
  B.~Beyguib, and A.~Poon, ``Wireless power transfer to deep-tissue
  microimplants,'' \emph{Proceedings of the National Academy of Sciences}, vol.
  111, no.~22, pp. 7974--7979, June 2014.

\bibitem{SEH1}
L.~Varshney, ``Transporting information and energy simultaneously,'' in
  \emph{Proc. IEEE Int. Symp. Inf. Theory}, July 2008, pp. 1612--1616.

\bibitem{SEH2}
P.~Grover and A.~Sahai, ``Shannon meets {T}esla: Wireless information and power
  transfer,'' in \emph{Proc. IEEE Int. Symp. Inf. Theory}, June 2010, pp.
  2363--2367.

\bibitem{Neg}
B.~Medepally and N.~Mehta, ``Voluntary energy harvesting relays and selection
  in cooperative wireless networks,'' \emph{{IEEE} Trans. Wireless Commun.},
  vol.~9, no.~11, pp. 3543--3553, Nov. 2010.

\bibitem{TSplit}
X.~Zhou, R.~Zhang, and C.~K. Ho, ``Wireless information and power transfer:
  Architecture design and rate-energy tradeoff,'' in \emph{Proc. IEEE Global
  Commun. Conf.}, Dec. 2012, pp. 3982--3987.

\bibitem{Zhou_AF}
A.~Nasir, X.~Zhou, S.~Durrani, and R.~Kennedy, ``Relaying protocols for
  wireless energy harvesting and information processing,'' \emph{{IEEE} Trans.
  Wireless Commun.}, vol.~12, no.~7, pp. 3622--3636, July 2013.

\bibitem{zhang2}
L.~Liu, R.~Zhang, and K.-C. Chua, ``Wireless information transfer with
  opportunistic energy harvesting,'' \emph{{IEEE} Trans. Wireless Commun.},
  vol.~12, no.~1, pp. 288--300, Jan. 2013.

\bibitem{dingzhiguo}
Z.~Ding, S.~Perlaza, I.~Esnaola, and H.~Poor, ``Power allocation strategies in
  energy harvesting wireless cooperative networks,'' \emph{{IEEE} Trans.
  Wireless Commun.}, vol.~13, no.~2, pp. 846--860, Feb. 2014.

\bibitem{GuICC14}
Y.~Gu and S.~Aissa, ``Interference aided energy harvesting in
  decode-and-forward relaying systems,'' in \emph{Proc. IEEE Int. Conf.
  Commun.}, June 2014, pp. 5378--5382.

\bibitem{R2}
Z.~Ding, I.~Krikidis, B.~Sharif, and H.~Poor, ``Wireless information and power
  transfer in cooperative networks with spatially random relays,'' \emph{{IEEE}
  Trans. Wireless Commun.}, vol.~13, no.~8, pp. 4440--4453, Aug. 2014.

\bibitem{R1}
H.~Chen, Y.~Li, Y.~Jiang, Y.~Ma, and B.~Vucetic, ``Distributed power splitting
  for swipt in relay interference channels using game theory,'' \emph{{IEEE}
  Trans. Wireless Commun.}, vol.~14, no.~1, pp. 410--420, Jan. 2015.

\bibitem{FreReUse}
J.~Du and Y.-C. Wu, ``Network-wide distributed carrier frequency offsets estimation and compensation via belief propagation,''
  \emph{{IEEE} Transactions on Signal Processing}, vol.~61, no.~23, pp. 5868--5877, 2013.
  
\bibitem{Multicell}
J.~Du and Y.-C. Wu, ``Distributed CFOs estimation and compensation in multi-cell cooperative networks,''
\emph{2013 International Conference on ICT Convergence (ICTC)}, pp. 117--121, 2013.
  

%\bibitem{Multicell}
%D.~Gesbert, S.~Hanly, H.~Huang, S.~Shamai~Shitz, O.~Simeone, and W.~Yu,
%  ``Multi-{C}ell {MIMO} cooperative networks: A new look at interference,''
%  \emph{{IEEE} J. Sel. Areas Commun.}, vol.~28, no.~9, pp. 1380--1408, Dec.
%  2010.

\bibitem{InterferAlignment}
V.~Cadambe and S.~Jafar, ``Interference alignment and degrees of freedom of the
  {K}-user interference channel,'' \emph{{IEEE} Trans. Inf. Theory}, vol.~54,
  no.~8, pp. 3425--3441, Aug. 2008.

\bibitem{YangGlobecom}
Y.~Yang, S.~Aissa, A.~Eltawil, and K.~Salama, ``An interference cancellation
  strategy for broadcast in hierarchical cell structure,'' in \emph{Proc. IEEE
  Global Commun. Conf.}, Dec. 2014, pp. 1792--1797.

\bibitem{Xia-Tcom12}
J.~Du and Y.-C.~Wu, ``Distributed clock skew and offset estimation in wireless sensor networks: asynchronous algorithm and convergence analysis,'' \emph{{IEEE} Trans. Wireless Communications}, vol.~12, no.~11, pp.
  5908--5917,  2013.
  
  
\bibitem{Gu}
Y.~Gu, S.~Ikki, and S.~Aissa, ``Opportunistic cooperative communication in the
  presence of co-channel interferences and outdated channel information,''
  \emph{{IEEE} Commun. Lett.}, vol.~17, no.~10, pp. 1948--1951, Oct. 2013.

\bibitem{Harvest-Use1}
O.~Ozel and S.~Ulukus, ``{AWGN} channel under time-varying amplitude
  constraints with causal information at the transmitter,'' in \emph{Proc. 45th
  Asilomar Conf. Signals, Syst. Comput.}, Nov. 2011, pp. 373--377.

\bibitem{BP}
J.~Du, S.~Ma, Y.-C.~Wu, Soummya Kar, and Jose MF Moura. "Convergence analysis of distributed inference with vector-valued Gaussian belief propagation." arXiv preprint arXiv:1611.02010 (2016).

\bibitem{Win2}
H.~Shin and M.~Win, ``{MIMO} diversity in the presence of double scattering,''
  \emph{{IEEE} Trans. Inf. Theory}, vol.~54, no.~7, pp. 2976--2996, July 2008.

\bibitem{Gradshteyn}
I.~S. Gradshteyn and I.~M. Ryzhik, \emph{Table of integrals, series and
  products}, 7th~ed.\hskip 1em plus 0.5em minus 0.4em\relax Academic Press,
  2007.

\bibitem{Majorization}
E.~Jorswieck and H.~Boche, \emph{Majorization and Matrix-Monotone Functions in
  Wireless Communications}.\hskip 1em plus 0.5em minus 0.4em\relax Foundations
  and Trends in Communication and Information Theory, 2007.

\end{thebibliography}
\end{document}